\pdfoutput=1
\documentclass[twocolumn]{aastex61}
\usepackage[space]{grffile}
\usepackage{amsfonts,amsmath,amssymb}
\usepackage{url}
\usepackage{latexsym}
\usepackage{hyperref} 
\usepackage{graphicx}
\usepackage{natbib}
\usepackage[utf8]{inputenc}
\usepackage{fancyref}
\usepackage{multirow}
\hypersetup{colorlinks=false,pdfborder={0 0 0},}

\newcommand{\rf}{\emph{realfast}}
\newcommand{\frb}{FRB 121102}

\begin{document}

\title{A Multi-telescope Campaign on \frb: Implications for the FRB Population}
\shorttitle{\frb\ Observing Campaign}
\shortauthors{Law et al.}

\author[0000-0002-4119-9963]{C.~J.~Law}
\affiliation{Department of Astronomy and Radio Astronomy Lab, University of California, Berkeley, CA 94720, USA}

\author{M.~W.~Abruzzo}
\affiliation{Haverford College, 370 Lancaster Ave, Haverford, PA 19041, USA}

\author{C.~G.~Bassa}
\affiliation{ASTRON, The Netherlands Institute for Radio Astronomy, Postbus 2, NL-7990 AA, Dwingeloo, The Netherlands}

\author{G.~C.~Bower}
\affiliation{Academia Sinica Institute of Astronomy and Astrophysics, 645 N. A'ohoku Place, Hilo, HI 96720, USA}

\author{S.~Burke-Spolaor}
\affiliation{National Radio Astronomy Observatory, Socorro, NM 87801, USA}
\affiliation{Department of Physics and Astronomy, West Virginia University, Morgantown, WV 26506, USA}
\affiliation{Center for Gravitational Waves and Cosmology, West Virginia University, Chestnut Ridge Research Building, Morgantown, WV 26505}

\author{B.~J.~Butler}
\affiliation{National Radio Astronomy Observatory, Socorro, NM 87801, USA}

\author{T.~Cantwell}
\affiliation{Jodrell Bank Centre for Astrophysics, Alan Turing Building, School of Physics \& Astronomy ,The University of Manchester, Oxford Road, Manchester M13 9PL, UK}

\author{S.~H.~Carey}
\affiliation{Astrophysics Group, Cavendish Laboratory, 19 J. J. Thomson Avenue, Cambridge CB3 0HE, UK}

\author{S.~Chatterjee}
\affiliation{Cornell Center for Astrophysics and Planetary Science and Department of Astronomy, Cornell University, Ithaca, NY 14853, USA}

\author{J.~M.~Cordes}
\affiliation{Cornell Center for Astrophysics and Planetary Science and Department of Astronomy, Cornell University, Ithaca, NY 14853, USA}

\author{P.~Demorest}
\affiliation{National Radio Astronomy Observatory, Socorro, NM 87801, USA}

\author{J.~Dowell}
\affiliation{Department of Physics and Astronomy, University of New Mexico, Albuquerque, NM 87131, USA}

\author{R.~Fender}
\affiliation{Centre for Astrophysical Surveys, University of Oxford, Denys Wilkinson Building, Keble Road, Oxford OX1 3RH, UK}

\author{K.~Gourdji}
\affiliation{Anton Pannekoek Institute for Astronomy, University of Amsterdam, Science Park 904, 1098 XH Amsterdam, The Netherlands}

\author{K.~Grainge}
\affiliation{Jodrell Bank Centre for Astrophysics, Alan Turing Building, School of Physics \& Astronomy ,The University of Manchester, Oxford Road, Manchester M13 9PL, UK}

\author{J.~W.~T.~Hessels}
\affiliation{ASTRON, The Netherlands Institute for Radio Astronomy, Postbus 2, NL-7990 AA, Dwingeloo, The Netherlands}
\affiliation{Anton Pannekoek Institute for Astronomy, University of Amsterdam, Science Park 904, 1098 XH Amsterdam, The Netherlands}

\author{J.~Hickish}
\affiliation{Department of Astronomy and Radio Astronomy Lab, University of California, Berkeley, CA 94720, USA}
\affiliation{Astrophysics Group, Cavendish Laboratory, 19 J. J. Thomson Avenue, Cambridge CB3 0HE, UK}

\author{V.~M.~Kaspi}
\affiliation{Department of Physics and McGill Space Institute, McGill University, 3600 University St., Montreal, QC H3A 2T8, Canada}

\author{T.~J.~W.~Lazio}
\affiliation{Jet Propulsion Laboratory, California Institute of Technology, Pasadena, CA 91109, USA}

\author{M.~A.~McLaughlin}
\affiliation{Department of Physics and Astronomy, West Virginia University, Morgantown, WV 26506, USA}
\affiliation{Center for Gravitational Waves and Cosmology, West Virginia University, Chestnut Ridge Research Building, Morgantown, WV 26505}

\author{D.~Michilli}
\affiliation{ASTRON, The Netherlands Institute for Radio Astronomy, Postbus 2, NL-7990 AA, Dwingeloo, The Netherlands}
\affiliation{Anton Pannekoek Institute for Astronomy, University of Amsterdam, Science Park 904, 1098 XH Amsterdam, The Netherlands}

\author{K.~Mooley}
\affiliation{Centre for Astrophysical Surveys, University of Oxford, Denys Wilkinson Building, Keble Road, Oxford OX1 3RH, UK}

\author{Y.~C.~Perrott}
\affiliation{Astrophysics Group, Cavendish Laboratory, 19 J. J. Thomson Avenue, Cambridge CB3 0HE, UK}

\author{S.~M.~Ransom}
\affiliation{National Radio Astronomy Observatory, Charlottesville, VA 22903, USA}

\author{N.~Razavi-Ghods}
\affiliation{Astrophysics Group, Cavendish Laboratory, 19 J. J. Thomson Avenue, Cambridge CB3 0HE, UK}

\author{M.~Rupen}
\affiliation{National Research Council of Canada, Herzberg Astronomy and Astrophysics, Dominion Radio Astrophysical Observatory, P.O. Box 248, Penticton, BC V2A 6J9, Canada}

\author{A.~Scaife}
\affiliation{Jodrell Bank Centre for Astrophysics, Alan Turing Building, School of Physics \& Astronomy ,The University of Manchester, Oxford Road, Manchester M13 9PL, UK}

\author{P.~Scott}
\affiliation{Astrophysics Group, Cavendish Laboratory, 19 J. J. Thomson Avenue, Cambridge CB3 0HE, UK}

\author{P.~Scholz}
\affiliation{National Research Council of Canada, Herzberg Astronomy and Astrophysics, Dominion Radio Astrophysical Observatory, P.O. Box 248, Penticton, BC V2A 6J9, Canada}

\author{A.~Seymour}
\affiliation{Arecibo Observatory, HC3 Box 53995, Arecibo, PR 00612, USA}

\author{L.~G.~Spitler}
\affiliation{Max-Planck-Institut f\"ur Radioastronomie, Auf dem H\"ugel 69, D-53121 Bonn, Germany}

\author{K.~Stovall}
\affiliation{National Radio Astronomy Observatory, Socorro, NM 87801, USA}
\affiliation{Department of Physics and Astronomy, University of New Mexico, Albuquerque, NM 87131, USA}

\author{S.~P.~Tendulkar}
\affiliation{Department of Physics and McGill Space Institute, McGill University, 3600 University St., Montreal, QC H3A 2T8, Canada}

\author{D.~Titterington}
\affiliation{Astrophysics Group, Cavendish Laboratory, 19 J. J. Thomson Avenue, Cambridge CB3 0HE, UK}

\author{R.~S.~Wharton}
\affiliation{Cornell Center for Astrophysics and Planetary Science and Department of Astronomy, Cornell University, Ithaca, NY 14853, USA}

\author{P.~K.~G.~Williams}
\affiliation{Harvard-Smithsonian Center for Astrophysics, Cambridge, MA, USA}

\begin{abstract}
We present results of the coordinated observing campaign that made the first subarcsecond localization of a Fast Radio Burst, \frb. During this campaign, we made the first simultaneous detection of an FRB burst by multiple telescopes: the VLA at 3~GHz and the Arecibo Observatory at 1.4~GHz. Of the nine bursts detected by the Very Large Array at 3~GHz, four had simultaneous observing coverage at other observatories. We use multi-observatory constraints and modeling of bursts seen only at 3~GHz to confirm earlier results showing that burst spectra are not well modeled by a power law. We find that burst spectra are characterized by a $\sim500$~MHz envelope and apparent radio energy as high as $10^{40}$\ erg. We measure significant changes in the apparent dispersion between bursts that can be attributed to frequency-dependent profiles or some other intrinsic burst structure that adds a systematic error to the estimate of DM by up to 1\%. We use \frb\ as a prototype of the FRB class to estimate a volumetric birth rate of FRB sources $R_{\rm{FRB}} \approx 5\times10^{-5}/N_r$\ Mpc$^{-3}$\ yr$^{-1}$, where $N_r$\ is the number of bursts per source over its lifetime. This rate is broadly consistent with models of FRBs from young pulsars or magnetars born in superluminous supernovae or long gamma-ray bursts, if the typical FRB repeats on the order of thousands of times during its lifetime.
\end{abstract}


\section{Introduction}

Fast Radio Bursts (FRBs) were discovered ten years ago with the detection of a millisecond-duration radio transient with an anomalously high dispersion measure \citep[DM;][]{2007Sci...318..777L}. The large DMs imply that FRBs originate outside of our Galaxy, potentially at cosmological distances, and are orders of magnitude more luminous than pulses from Galactic pulsars \citep{2013Sci...341...53T}. There are now 22 FRBs publicly known \footnote{See \url{http://www.astronomy.swin.edu.au/pulsar/frbcat} \citep{2016PASA...33...45P}.} with DMs as high as 1600 pc cm$^{-3}$ and temporal widths of order milliseconds. Both their energetics and distance have inspired a wide variety of models and astrophysical applications \citep[e.g.,][]{2014ApJ...780L..33M, 2014ApJ...797...70K, 2015MNRAS.450L..71F, 2016MNRAS.458L..19C, 2016MNRAS.457..232C, 2016MNRAS.462L..16P}. However, that potential was limited by the lack of a definitive association of an FRB to an extragalactic host.

This paper is part of a series based on the first localization of an FRB and its unambiguous association to an extragalactic host \citep{LOC, OPT, EVN}. \frb, also known as the ``repeating FRB'', was discovered \citep{2014ApJ...790..101S} in data acquired in 2012 as part of the PALFA survey of the Arecibo Observatory \citep{2006ApJ...637..446C, 2015ApJ...812...81L}. In mid 2015, new Arecibo observations revealed a series of bursts at a similar DM and sky position that rules out cataclysmic models for this source \citep{2016Natur.531..202S}. The typical DM measured by early Arecibo observations was 559 pc cm$^{-3}$ \citep{2016arXiv160308880S}, although somewhat higher values have been seen in more recent observations (560.5 pc cm$^{-3}$; Hessels et al., in prep).

Beginning in August 2016 (MJD 57623), we made the first of nine detections of \frb\ with the Karl G.\ Jansky Very Large Array \citep[VLA;][]{LOC} and localized it with a precision of 0\farcs1. Deep optical observations with the Gemini Observatory associated the FRB with a host galaxy at $z=0.193$\ \citep{OPT} and the European VLBI Network detected four more bursts to localize the source with a precision of 0.01\arcsec, four orders of magnitude better than any other FRB \citep[precision of $\sim40$\ pc in linear distance;][]{EVN}. The lookback and luminosity distances for \frb\ are 746 and 972 Mpc, respectively, assuming a concordance cosmology with parameters given by \citet{2016A&A...594A..13P}. If \frb\ is representative of all FRBs, then we should expect them to be useful as probes of the intergalactic medium and their host galaxies \citep{2015MNRAS.451.3278M}. Thus, the confirmation of a cosmological distance for \frb\ has begun to fulfil the promise implied by the first FRB detection.

Many new models for the origin of FRBs have been developed in response to the localization of \frb\ \citep{2017arXiv170104815K, 2017arXiv170102370M, 2017arXiv170104094Z, 2017arXiv170102492D, 2017arXiv170208644B, 2017arXiv170300393T}. The association of \frb\ with a compact, persistent radio source is consistent with bursts coming from a young magnetar that powers a luminous pulsar wind nebula \citep{2017arXiv170104815K}. At the same time, \frb\ is found in a low-metallicity dwarf galaxy \citep{OPT}, which was not predicted by origin models that scale with stellar mass or star formation \citep{2017arXiv170400022N}. These galaxies are the preferred environment for long GRBs and hydrogen-poor superluminous supernovae (LGRB and SLSN-I, respectively), which have been suggested are signatures of magnetar birth \citep{2008AJ....135.1136M, 2014ApJ...787..138L}.

This paper presents an analysis of the spectral properties of VLA bursts implied by simultaneous observing at Arecibo, Effelsberg, the first station of the Long-Wavelength Array \citep[LWA1; ][]{2013ITAP...61.2540E}, and the Arcminute Microkelvin Imager Large Array \citep[AMI-LA; ][]{2008MNRAS.391.1545Z}. This includes the first simultaneous detection of an FRB at two observatories, between the VLA and Arecibo. If we assume that \frb\ is representative of all FRBs, we can use it to constrain the physical processes at play in the overall FRB population. The repetition of \frb\ also has strong implications for calculations of their rate of occurrence \citep{2016MNRAS.458L..89C} and comparison to other classes of transient, such as superluminous supernovae \citep{OPT}.

In Section \ref{sec:obs}, we describe the multi-telescope observing campaign and a refined analysis of the nine VLA bursts. Section \ref{sec:res} presents the first spectrum of an FRB simultaneously detected at multiple telescopes, confirming that burst spectra cannot be modeled with a single spectral index. We then model the dynamic spectra to characterize the burst spectra, energies, and dispersion properties. Section \ref{sec:disc} discusses the properties of \frb\ bursts and their impact on inferences about the birth rate of FRB sources and strategies to find and/or localize new FRBs.

\section{Observations}
\label{sec:obs}

\begin{figure*}[t]
\begin{center}
\includegraphics[width=2\columnwidth]{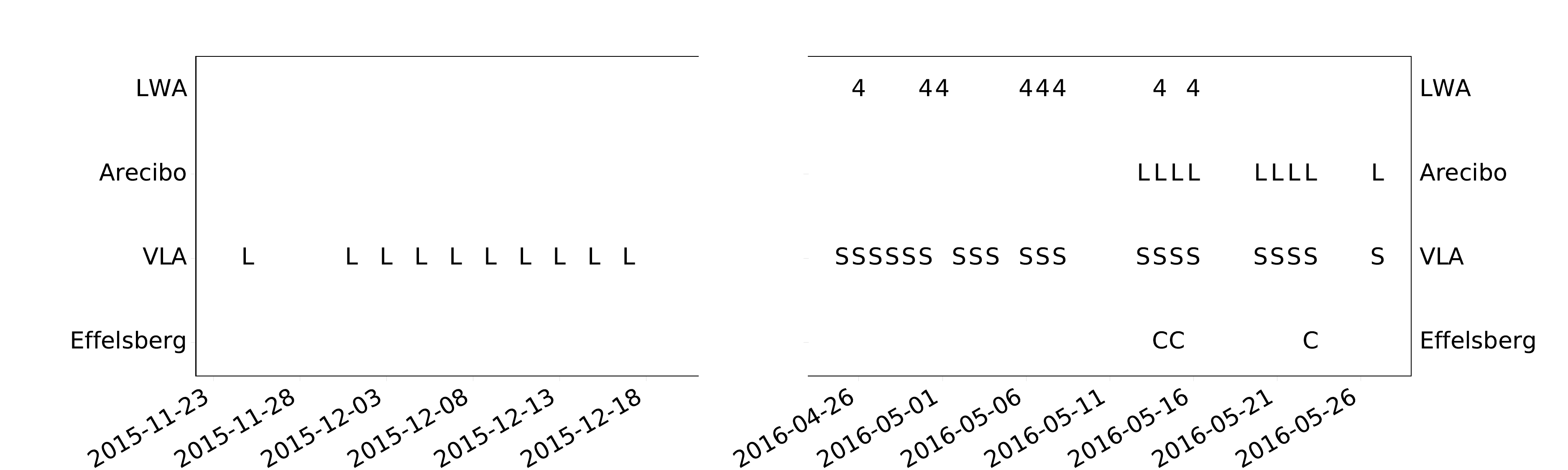}

\includegraphics[width=2\columnwidth]{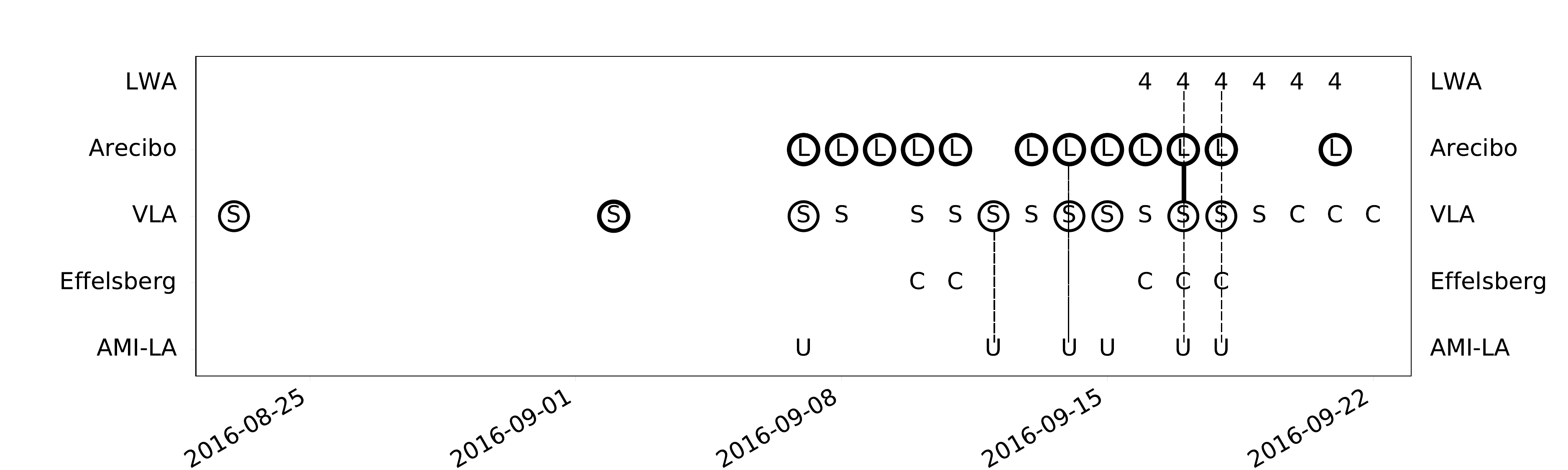}
\caption{The top and bottom panels summarize the observations and detections of \frb\ in the 2015 and 2016 campaigns. Circles highlight observations that detected bursts from \frb\ with multiple burst detections indicated with a heavy line. The black dashed lines show the VLA burst detections with simultaneous coverage at other telescopes and the solid line shows the simultaneous burst detection. Circled observations with no line indicate bursts that did not have simultaneous observing coverage at the VLA. Days with observations are indicated with radio band designations 4, L, S, C, U referring to radio frequencies of 70~MHz, 1.4, 3, 4.5, and 15~GHz, respectively.
\label{fig:sched}}
\end{center}
\end{figure*}

The data presented here were obtained from multiple programs and telescopes with a goal of interferometrically localizing \frb\ with the VLA. We coordinated observing between the VLA, Arecibo, Effelsberg, LWA1, and AMI-LA telescopes, as shown in Figure \ref{fig:sched} and Table \ref{tab:obs}. Below, we summarize these observations, with a focus on those conducted simultaneously with VLA burst detections from \frb.

Computational (Jupyter) notebooks to reproduce the transient detection, localization, and analysis presented here can be found at \url{https://github.com/caseyjlaw/FRB121102}. Time cut-out visibility data and calibration products are available at \url{https://doi.org/10.7910/DVN/TLDKXG}. Original VLA visibility data are available under NRAO program codes 16A-459 and 16A-496 and can be downloaded at \url{http://archive.nrao.edu}.

\startlongtable
\begin{deluxetable}{lcccc}
\small
\tablecaption{Table of VLA Observations \label{tab:obs}}
%\centering
%\begin{tabular}{lcccccc}
\startdata
MJD & Date/time & Duration & Freq. & Bursts \\
 & (ymd/hms) & (min) & (GHz) & \\ \hline
57351 & 2015-11-25/3:25:29 & 60 & 1.4 & 0 \\
57357 & 2015-12-01/5:31:31 & 60 & 1.4 & 0 \\
57359 & 2015-12-03/2:53:57 & 60 & 1.4 & 0 \\
57361 & 2015-12-05/4:45:44 & 60 & 1.4 & 0 \\
57363 & 2015-12-07/4:37:50 & 60 & 1.4 & 0 \\
57365 & 2015-12-09/9:29:25 & 60 & 1.4 & 0 \\
57367 & 2015-12-11/9:22:27 & 60 & 1.4 & 0 \\
57369 & 2015-12-13/9:13:7 & 60 & 1.4 & 0 \\
57371 & 2015-12-15/9:6:16 & 60 & 1.4 & 0 \\
57373 & 2015-12-17/8:57:51 & 60 & 1.4 & 0 \\
57503 & 2016-04-25/23:21:39 & 240 & 3 & 0  \\
57504 & 2016-04-26/23:14:31 & 120 & 3 & 0  \\
57505 & 2016-04-27/23:26:09 & 120 & 3 & 0  \\
57506 & 2016-04-28/22:41:31 & 120 & 3 & 0  \\
57507 & 2016-04-29/22:37:34 & 120 & 3 & 0  \\
57508 & 2016-04-30/19:14:46 & 60 & 3 & 0  \\
57510 & 2016-05-02/23:14:18 & 60 & 3 & 0 \\
57511 & 2016-05-03/22:53:26 & 60 & 3 & 0  \\
57514 & 2016-05-06/18:25:12 & 60 & 3 & 0  \\
57515 & 2016-05-07/19:34:13 & 60 & 3 & 0  \\
57516 & 2016-05-08/19:30:20 & 60 & 3 & 0  \\
57521 & 2016-05-13/17:12:48 & 135 & 3 & 0  \\
57522 & 2016-05-14/17:14:11 & 135 & 3 & 0  \\
57523 & 2016-05-15/17:15:26 & 135 & 3 & 0  \\
57524 & 2016-05-16/16:59:52 & 135 & 3 & 0  \\
57528 & 2016-05-20/16:45:46 & 135 & 3 & 0  \\
57529 & 2016-05-21/16:44:6 & 135 & 3 & 0  \\
57530 & 2016-05-22/16:43:53 & 135 & 3 & 0  \\
57531 & 2016-05-23/16:44:11 & 135 & 3 & 0  \\
57535 & 2016-05-27/16:29:34 & 135 & 3 & 0  \\
57623 & 2016-08-23/17:26:28 & 54 & 3 & 1  \\
57633 & 2016-09-02/15:52:17 & 54 & 3 & 2  \\
57638 & 2016-09-07/10:14:50 & 120 & 3 & 1  \\
57639 & 2016-09-08/10:14:40 & 120 & 3 & 0  \\
57641 & 2016-09-10/9:59:36 & 120 & 3 & 0  \\
57642 & 2016-09-11/9:59:49 & 120 & 3 & 0  \\
57643 & 2016-09-12/9:15:19 & 120 & 3 & 1\tablenotemark{d} \\ 
57644 & 2016-09-13/9:23:59 & 120 & 3 & 0  \\
57645 & 2016-09-14/9:20:23 & 120 & 3 & 1\tablenotemark{a,d} \\
57646 & 2016-09-15/9:16:29 & 120 & 3 & 1  \\
57647 & 2016-09-16/9:11:23 & 120 & 3 & 0  \\
57648 & 2016-09-17/8:59:20 & 120 & 3 & 1\tablenotemark{a,*,b,c,d} \\ 
57649 & 2016-09-18/8:59:27 & 120 & 3 & 1\tablenotemark{a,b,c,d} \\
57650 & 2016-09-19/8:44:32 & 120 & 3 & 0  \\
57651 & 2016-09-20/8:44:33 & 120 & 3 & 0  \\
57651 & 2016-09-20/17:19:3 & 120 & 6 & 0  \\
57652 & 2016-09-21/9:13:57 & 120 & 6 & 0  \\
57653 & 2016-09-22/9:12:24 & 120 & 6 & 0  \\ \hline
\enddata
\tablecomments{Times are in UT. Duration includes overhead, which is typically 15\%. Frequency is the approximate center of the bandwidth.}
\tablenotetext{a}{Arecibo coverage at 1.4~GHz.}
\tablenotetext{*}{Arecibo detection at 1.4~GHz.}
\tablenotetext{b}{Effelsberg coverage at 4.5~GHz.}
\tablenotetext{c}{LWA1 coverage near 62 and 78~MHz.}
\tablenotetext{d}{AMI-LA coverage at 15~GHz.}
%\end{tabular}
%\tablenotetext{a}{}
%\label{tab:obs}
\end{deluxetable} 

\subsection{VLA}

The \frb\ observing campaign started in November 2015 with a 10~hr campaign ($\sim1$~hr per session) observed at 1.4~GHz in the compact D configuration. In April through May 2016, we conducted a 40~hr campaign ($\sim2$~hr per session) at 3~GHz in the C and CnB configurations in coordination with Arecibo \citep{2016arXiv160308880S}. We concluded with a 40~hr, coordinated campaign ($\sim2$~hr per session) from August through September 2016 in the B configuration and during the move to the most extended A configuration. In the late-2016 campaign, the first 34 hours of VLA observations were made at 3~GHz, while the last 6 hours were observed at 6~GHz. These observing session times are inclusive of calibration and overhead, which typically amounts to 15\% of the total observing time.

All VLA fast-sampled data were observed with 5~ms sampling, 256 channels, and dual-circular polarization \citep{2015ApJ...807...16L}. The total bandwidths at L (1.4~GHz), S (3~GHz), and C (6~GHz) bands were 256, 1024, and 2048~MHz, respectively, corresponding to channel bandwidths of  1, 4, and 8~MHz. The dispersion smearing across a channel is $\sim1.7$, 0.7, and 0.2~ms in the same respective bands. The 3~GHz data were recorded in eight spectral windows with 32 channels each and had a sensitivity of 5~mJy in 5~ms ($1\sigma$).

Observations in August and September were searched by a prototype version of \rf\footnote{See \url{http://realfast.io} \citep{2017AAS...22933002L}.}. \rf\ is a real-time, fast imaging transient search system. The current prototype runs on CPU-based hardware that is normally dedicated to the VLA correlator; for this experiment, it runs the transient search pipeline software called \emph{rtpipe}\footnote{See \url{https://github.com/caseyjlaw/rtpipe} \citep{2015ApJ...807...16L}.}. Images were formed for each integration with DMs of 0, 546, 556.9, 560, and 565 pc cm$^{-3}$ and a time resampling grid of 5, 10, 20, 40, and 80 ms. This DM grid was chosen to maintain 90\% sensitivity to the nominal DM range of 540--570 pc cm$^{-3}$. Gain calibration was made from observations of J0555+3948 by an automated system (telcal), which uses phase-only calibration. A flux scale was calculated for each spectral window from an observation of 3C48 and applied to all burst spectra and has an accuracy of about 10\%.

Burst detections and localizations were made within 5--10 hours of data being recorded. The transient search starts when data are recorded and proceeds slower than real-time, so we refer to it as ``quasi real-time''. For each trial DM, integration, and time scale, we form an image and calculate the S/N ratio for the peak pixel in the dirty image. We empirically identified S/N thresholds of 6.4 and 7.4 as useful to capture data quality statistics and candidates for inspection, respectively. The higher threshold is relatively unlikely to be triggered by thermal noise in this configuration, so \rf\ generates a more detailed (and computationally intensive) candidate visualization that includes an image and spectrum. All visibilities are recorded so detailed analysis, including improved calibration and localization, can be conducted offline. 

\subsection{Arecibo}

Arecibo observed with the L-wide receiver using the PUPPI pulsar backend. The observational frequency range was 1.15 to 1.73~GHz and frequency resolution was 1.5625~MHz, which has a typical sensitivity of 2~mJy in 2~ms ($1\sigma$). We recorded intnsities of the each of the two orthogonal linear polarization signals and their cross product. Full Stokes polarization intensity spectra can then be generated with a time resolution of 10.24~$\mu$s. Each frequency channel was coherently dedispersed to 557~pc cm$^{-3}$, significantly reducing intra-channel dispersion smearing. The full width at half maximum (FWHM) beam size at band center is 3.3\arcmin.

In total, 11 Arecibo observations had some simultaneous coverage with the VLA. All of these observations were conducted after the VLA localization, so they were pointing directly at \frb. Three of those observations had simultaneous coverage of VLA bursts and one of those Arecibo observations detected the VLA burst. 
% During the first VLA burst with Arecibo coverage (MJD 57643), the PUPPI 1.4~GHz recording system failed so data were recorded at C band with the mock spectrometer. The C-band observations were recorded at frequencies from 4078 to 4245~MHz and had a sensitivity of 3.6~mJy in 2~ms ($1\sigma$). No detection was made in those Arecibo data. % levels set wrong. useless c-band ao data.
Overall, there were many more bursts detected at Arecibo than with the VLA, including some Arecibo bursts with simultaneous VLA upper limits. A more detailed analysis of those bursts will be presented elsewhere (Michilli et al, in prep).

\subsection{Effelsberg}

Effelsberg observations were conducted with the S60mm receiver and recorded total intensity spectra with the PFFTS backend in pulsar search mode. The observations had a time resolution of 65.5~$\mu$s and 128 frequency channels. The receiver has a system equivalent flux density of 18 Jy and a FWHM beam size of 2.4\arcmin\ at 4.85~GHz. 

Five Effelsberg observations were made pointing at the known location of \frb\ and had some simultaneous coverage with the VLA. Two of these observations were simultaneous with VLA bursts. The nominal receiver bandwidth is from 4.6 to 5.1~GHz, but a configuration error reduced effective bandwidth to 100~MHz centered at 4.85~GHz for both of these observations during VLA bursts. The sensitivity in these observations was about 28 mJy in 2 ms ($1\sigma$), which is two times worse than the full-bandwidth value. No burst was detected in either of the two observations that were simultaneous with a VLA burst detection.

\subsection{LWA1}

Beginning in April 2016, VLA observations of \frb\ were simultaneously observed with LWA1 when possible.  The LWA1 observations were automatically scheduled through the Heuristic Automation for LWA1 (HAL) system.  Briefly, the HAL system receives messages via the internet that signals the start of a VLA observation. If the source is visible and there are no high priority LWA1 observations scheduled, the HAL system automatically generates an observing schedule, re-configures the telescope, and alerts the observatory staff of the change. The time delay from HAL receiving a message to starting an observation is typically one to two minutes.

The LWA1 observations consisted of a single phased-array beam centered on the location of \frb\ with two 4096-channel spectral windows spanning frequencies from 52.2--71.8~MHz and 68.2--87.8~MHz. The observations were taken in a spectrometer mode with a channel size of 4.8~kHz and a sample time of 160~ms. The integration time was set to be equal to the dispersion smearing across a single channel at 50~MHz for DM$\approx560$\ pc cm$^{-3}$. 

Two of the VLA detected pulses, on MJD 57648 and 57649, had simultaneous coverage with LWA1. We use the modeled LWA1 antenna sensitivity as a function of zenith angle to estimate an system equivalent flux density of 9.5 kJy and 8.7 kJy for the two bursts, respectively. The data were de-dispersed into time series with DMs ranging from 500 to 600 pc cm$^{-3}$\ using a step size of 1.0 pc cm$^{-3}$. Each spectral window was searched using PRESTO \citep{2001PhDT.......123R} with a boxcar matched-filtering width from the native time resolution up to 48 seconds. We also visually inspected the dedispersed time series for the DM range 557 to 560 pc cm$^{-3}$.

No dispersed pulses were found with significance greater than $10\sigma$, equivalent to a flux density limit of $\sim60$\ Jy for a width of 160~ms. The system equivalent flux density does not change much between the two windows. The effective sensitivity to a typical 2~ms pulse width is 4.8~kJy.

\subsection{AMI-LA}

We observed \frb\ with AMI-LA for 3 hours each on six epochs starting at MJDs 57638, 57643, 57645, 57646, 57648, and 57649. Observations were made with the new digital correlator having 4096 channels across a 5~GHz bandwidth between 13--18~GHz with a 1~s integration time. The phase calibrator, J0518+3306, was observed every 12 minutes for about 1.5 minutes. The AMI-LA data were binned to eight 0.625~GHz channels and processed (interference excision and calibration) with a fully-automated pipeline, AMI-REDUCE \citep[e.g.,][]{2013MNRAS.429.3330P}. Daily measurements of 3C48 and 3C286 were used for the absolute flux calibration, which is good to about 10\%. 

We inspected the calibrated visibilities, and did not find any signal above 30 mJy in the 1~s samples at and in the vicinity of the detected bursts. This corresponds to a sensitivity of 15~Jy for an assumed pulse width of 2~ms. Concatenating and imaging the 12 hours of calibrated data with the CASA tasks {\it concat} and {\it clean} also does not yield any significant detection at the FRB location. Although the statistical $3\sigma$\ upper limit is 60~$\mu$Jy, extended mJy-level sources in the field cause sidelobe confusion (the AMI-LA angular resolution is $\sim$30\arcsec), and the actual upper limit is larger. We introduced artificial point sources at the FRB location using the CASA {\it sm} tool, and found that these sources can be recovered at $3\sigma$\ as long as their peak flux densities are more than $\sim100~\mu$Jy. Hence, we place an upper limit of $100\pm10~\mu$Jy on any quiescent or possible radio flaring on $\sim$days timescales from the FRB. This limit is similar to the flux density measured by the VLA at 12~GHz for the persistent source \citep{LOC}.

\section{Results}
\label{sec:res}

\subsection{Multi-Observatory Burst Spectrum}

\begin{figure}[htb]
\begin{center}
 \includegraphics[trim=250 400 250 80, clip, width=\columnwidth]{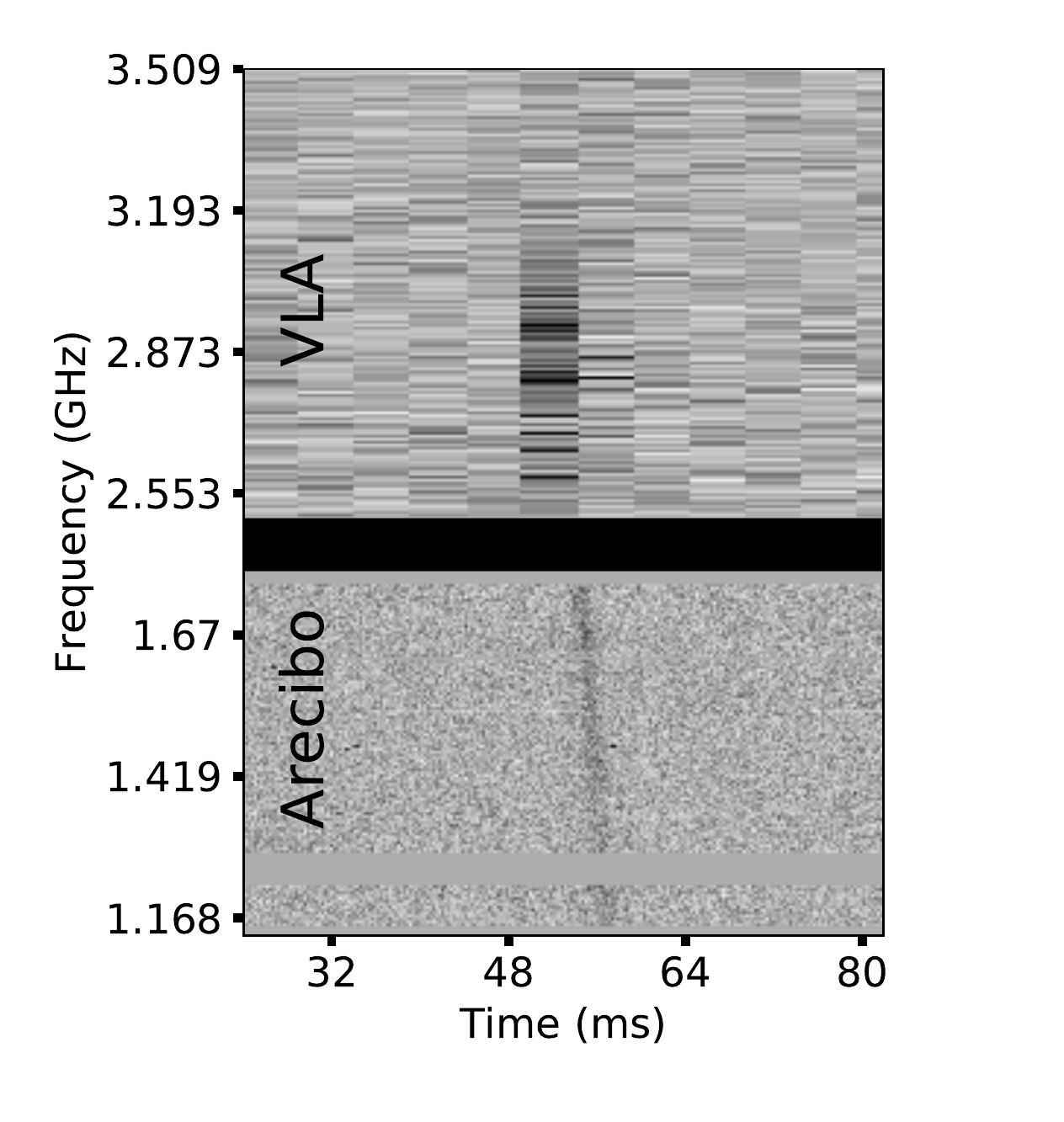}
 \caption{Composite dynamic spectrum for the burst from \frb\ on MJD 57648 with data from the VLA and Arecibo observatories. The time axis is measured in milliseconds relative to MJD 57648.437788874842 with times corrected to the barycenter and infinite frequency (assuming DM$=560.5$\ pc $^{-3}$). The thick black line separates data from the VLA (2.5 to 3.5~GHz) and Arecibo (1.1 to 1.8~GHz) and flagged data are grey. For clarity, the VLA and Arecibo data are independently normalized to units of S/N per pixel; Arecibo is 5 times more sensitive than the VLA.
 \label{fig:sgram}}
\end{center}
\end{figure}

Figure \ref{fig:sgram} shows a dynamic spectrum for the first FRB burst to be detected simultaneously at two telescopes. This burst, on MJD 57648, was detected both by the VLA and Arecibo while upper limits were measured by simultaneous observations at the other three telescopes. In total, four of the VLA bursts had simultaneous observing coverage with either Arecibo, Effelsberg, LWA1, or AMI-LA. Two of the VLA bursts had simultaneous observing coverage by all four observatories, including the burst detected by Arecibo.

To generate the multi-telescope dynamic spectrum, the VLA and Arecibo data were resampled to the same temporal grid relative to the start time of the Arecibo data. We corrected for barycentric and dispersion measure offsets by assuming a DM of 560.5 pc cm$^{-3}$ (Hessels et al., in prep). The dispersion delay was 189 and 734~ms from infinite frequency to the top of the VLA and Arecibo bands, respectively. An error in the DM of 1 pc cm$^{-3}$\ corresponds to a delay time correction error of 0.3 and 1.4~ms for the VLA and Arecibo bands, respectively. We used PRESTO to calculate a relative barycentric time correction of 3.8~ms between the VLA and Arecibo at the time of observation.

The regridded dynamic spectrum has an apparent DM error that is evident both as a frequency-dependent time drift within the Arecibo observation and as an offset between the Arecibo and VLA bursts. Both of these drifts are consistent with an apparent DM of $\sim$565 pc cm$^{-3}$. More refined modeling of the VLA dynamic spectrum alone (\S \ref{sec:spec} and Table \ref{tab:spec}) are consistent with this higher DM value.

\begin{figure}[htb]
\begin{center}
 \includegraphics[width=\columnwidth]{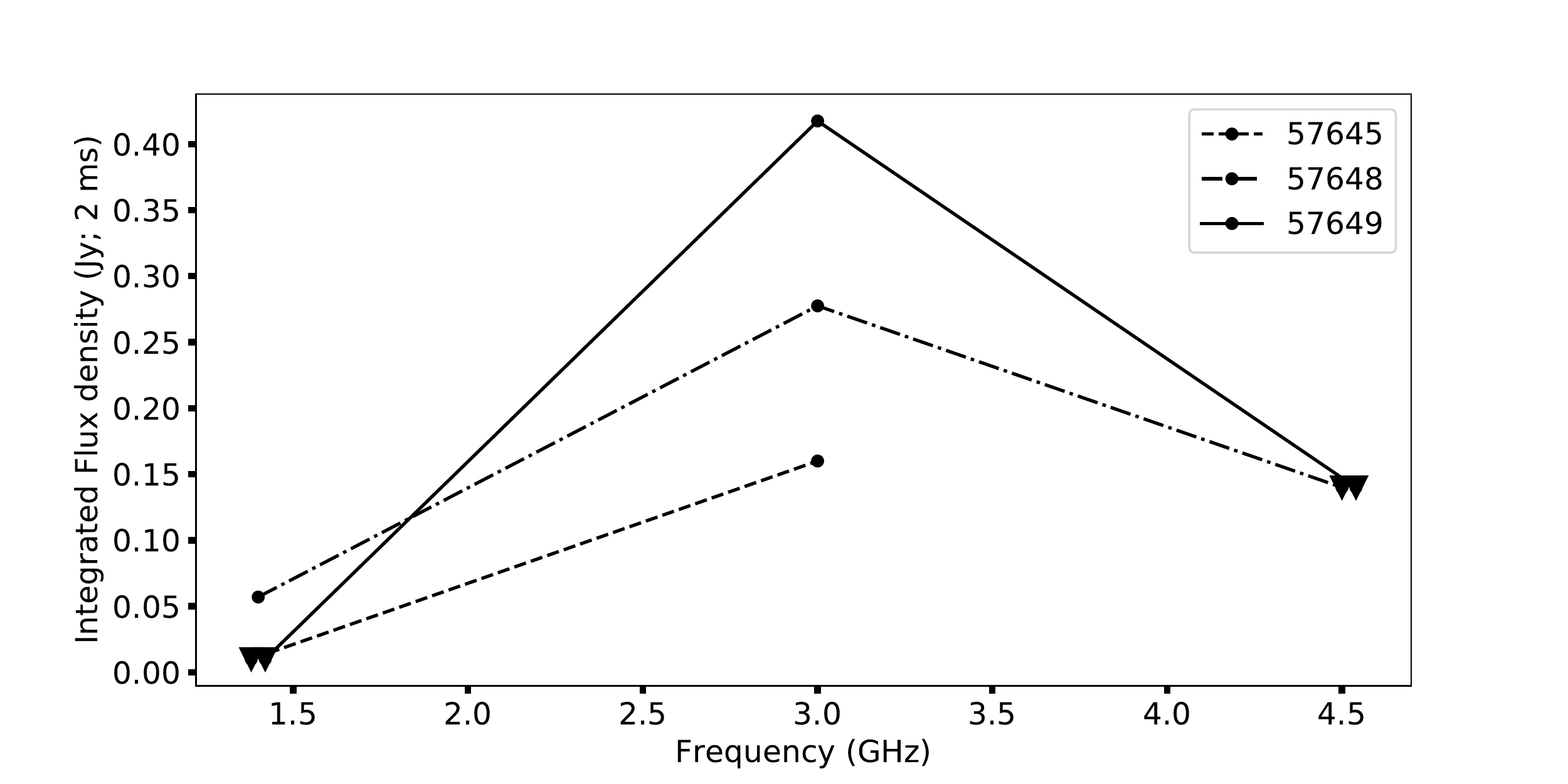}
 \caption{Broadband spectral measurements and limits for three bursts of \frb\ with observing coverage by Arecibo or Effelsberg. Upper limits assume a pulse width of 2~ms and a $5\sigma$ detection threshold. Measurements (dots) and limits (triangles) are shown for Arecibo, VLA, and Effelsberg at 1.4, 3, and 4.5~GHz, respectively; limits from LWA1 and AMI-LA are not visible on this scale. Errors in flux density are comparable to the symbol size and are not shown. Overlapping points are offset by tens of MHz in frequency for clarity.
 \label{fig:multi}}
\end{center}
\end{figure}

Figure \ref{fig:multi} shows coarse spectra built from integrated flux densities measured (or limited) for the three VLA bursts with observing coverage by Arecibo or Effelsberg. The burst on MJD 57648 was detected with VLA and Arecibo with significances of 25$\sigma$\ and 39$\sigma$, which corresponds to peak flux densities in 5~ms of $111\pm5$\ mJy and $14\pm0.4$\ mJy at 3 and 1.4~GHz, respectively. The VLA observations under-resolve the pulses in time, so to compare these values, we assume a typical pulse width of 2~ms. Under this assumption, the VLA and Arecibo integrated flux densities are $278\pm13$\ and $57\pm2$\ mJy at 3 and 1.4~GHz, respectively. Assuming a power law flux density model ($S_{\nu} \propto \nu^{\alpha}$), we find a spectral index $\alpha=2.1$. This is inconsistent with the spectral index limit implied by the Effelsberg nondetection at 4.5~GHz ($\alpha<-1.7$\ for a 2~ms burst and $5\sigma$ limit).

Overall, the bursts with simultaneous observing coverage are not well described by a power-law model. Two Arecibo nondetections and two Effelsberg nondetections of VLA bursts place mutually inconsistent lower and upper limits on the spectral index. The Arecibo and Effelsberg upper limits at 1.4 and 4.5~GHz, respectively, limit all burst spectral indices $\alpha_{1.4/3}>+4.9$ and $\alpha_{3/4.5}<-2.7$. These are consistent with limits from the LWA1 and AMI-LA, which require $\alpha_{0.07/3}>-2.4$\ and $\alpha_{3/15}<1.5$ for bursts on MJD 57643 and 57649 (the brightest VLA bursts with observing coverage), respectively. The strictest limits on $\alpha_{1.4/3}$\ and $\alpha_{3/4.5}$\ are both derived from the burst on MJD 57649, which shows that a power law model is inappropriate even for a single burst.

\subsection{VLA Bursts}

\subsubsection{Spectrotemporal Modeling}
\label{sec:spec}

This paper refines the analysis of the nine VLA radio burst spectra described in \citet{LOC} in a few ways. We use a better calibration scheme and have optimized the detection significance over a fine grid of DM ($\Delta \rm{DM}=1\ \rm{pc}\ \rm{cm}^{-3}$). After calibration and flagging, the visibility phases were rotated to the best-fit location \citep[RA, Dec $=$ 05h31m58.70s, +33d08m52.5s;][]{LOC} to extract a Stokes I spectrum that maximizes the image S/N for each burst.

Table \ref{tab:spec} summarizes the spectrotemporal properties of all nine VLA bursts. Parameters such as integrated flux density, S/N, and Stokes V were measured from the burst properties integrated in frequency. The circular polarization fraction (Stokes V/I) was estimated as $(RR-LL)/(RR+LL)\approx3\%$, which is comparable to the systematic error expected from beam squint at this location in the primary beam \footnote{See EVLA memo \#195 at \url{https://library.nrao.edu/public/memos/evla/EVLAM_195.pdf}.}. Given that systematic effects dominate the apparent circular polarization, this observation limits the fractional circular polarization to less than 3\%.

\begin{table*}
\caption{Properties of \frb\ Bursts Seen by VLA at 3~GHz}
%\centering
\begin{tabular}{cc|rrr|ccrrrr}
\multicolumn{2}{c|}{Time} & \multicolumn{3}{c|}{Observed properties} & \multicolumn{6}{c}{Modeled properties} \\ \hline
Calendar day & Burst time   & S/N & $S_{\rm{I, 5~ms}}$	& $S_{\rm{V}}$ 	& DM 			& W 			& $S_{\rm{I,peak}}$ & Center & FWHM & $E_{\rm{int}}$ \\
(2016)       &  (MJD)     &   (image)    & (mJy) 			& (mJy) 		    & (pc cm$^{-3})$ 	& (ms) 			& (mJy) 			& (GHz)  & (MHz) & ($10^{38}$\ erg)\\ \hline
23 Aug & 57623.74402686      & 38 		& 194			& +3				& 567$\pm2$ 		& 2.0$\pm0.2$	 		& 690 		& 2.8 		& 290 & 12 	\\
2 Sep & 57633.67986367      & 179 		& 1500			& --35 				& 568.2$\pm0.2$ 	& 2.05$\pm0.02$			& 3340 		& 3.2 		& 510 & 98		\\
2 Sep\tablenotemark{c} & 57633.69515938 & 15 & 69		& +2				& 562$^{+4}_{-6}$ 	& 2.5$^{+0.9}_{-0.6}$	& $>$430 	& $<$2.5	& $<$290 & 7 	\\
7 Sep & 57638.49937435      & 12 		& 55			& +5 				& 567$^{+7}_{-9}$ 	& 1.3$^{+1.4}_{-0.8}$	& 130 		& 3.1 		& 420 & 3  	\\
12 Sep\tablenotemark{a} & 57643.45730263  & 100 		& 508 & --5 		& 565.6$^{+0.6}_{-0.5}$ & 1.9$\pm0.1$		& 1170 		& 2.8 		& 510 & 34 		\\
14 Sep\tablenotemark{a} & 57645.42958602   & 13 		& 64 & +3			& 563$^{+5}_{-4}$ 	& 1.1$\pm0.7$		 	& 170 		& 2.8 		& 380 & 4  	\\
15 Sep\tablenotemark{c} & 57646.46600650 & 20 & 87		& +1				& 569$\pm5$ 		& 2.5$^{+0.9}_{-1.4}$	& $>$420 	& $<$2.5 	& $<$430 & 10 	\\	
17 Sep\tablenotemark{a,b} & 57648.43691490 & 25 & 111	& +9				& 564$\pm2$ 		& 1.4$^{+0.3}_{-0.4}$	& 260 		& 2.8 		& 470 & 7  	\\
18 Sep\tablenotemark{a} & 57649.45175697   & 36 		& 167 & +1			& 567$\pm2$ 		& 2.1$\pm0.5$		 	& 290 		& 3.0 		& 690 & 12 	\\ \hline
\end{tabular}
\tablecomments{Burst times are topocentric at the VLA at 3.5~GHz. $S_{\rm{V}}$ is the measured circular polarization of the burst, which dominated by systematic effects. All error ranges represent 68\% confidence intervals.}
\tablenotetext{a}{Simultaneously observing coverage with either Arecibo, Effelsberg, LWA1, or AMI-LA.}
\tablenotetext{b}{Detected simultaneously with Arecibo.}
\tablenotetext{c}{Best-fit Gaussian is not centered in 3~GHz band, so spectral parameters are limits.}
\label{tab:spec}
\end{table*} 

\begin{figure*}[ht]
\begin{center}
 \begin{minipage}{2\columnwidth}
  \includegraphics[width=0.5\columnwidth]{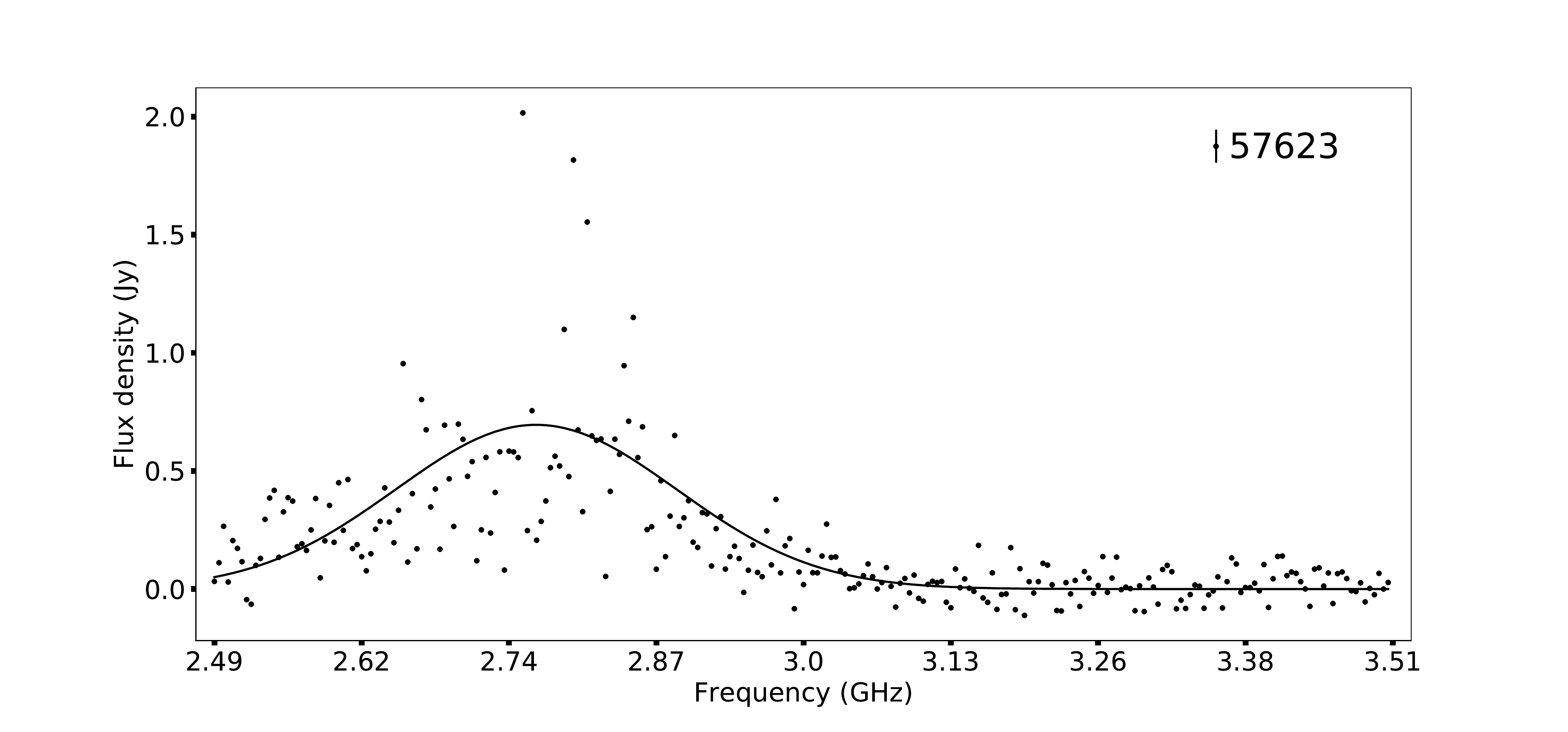}
  \includegraphics[width=0.5\columnwidth]{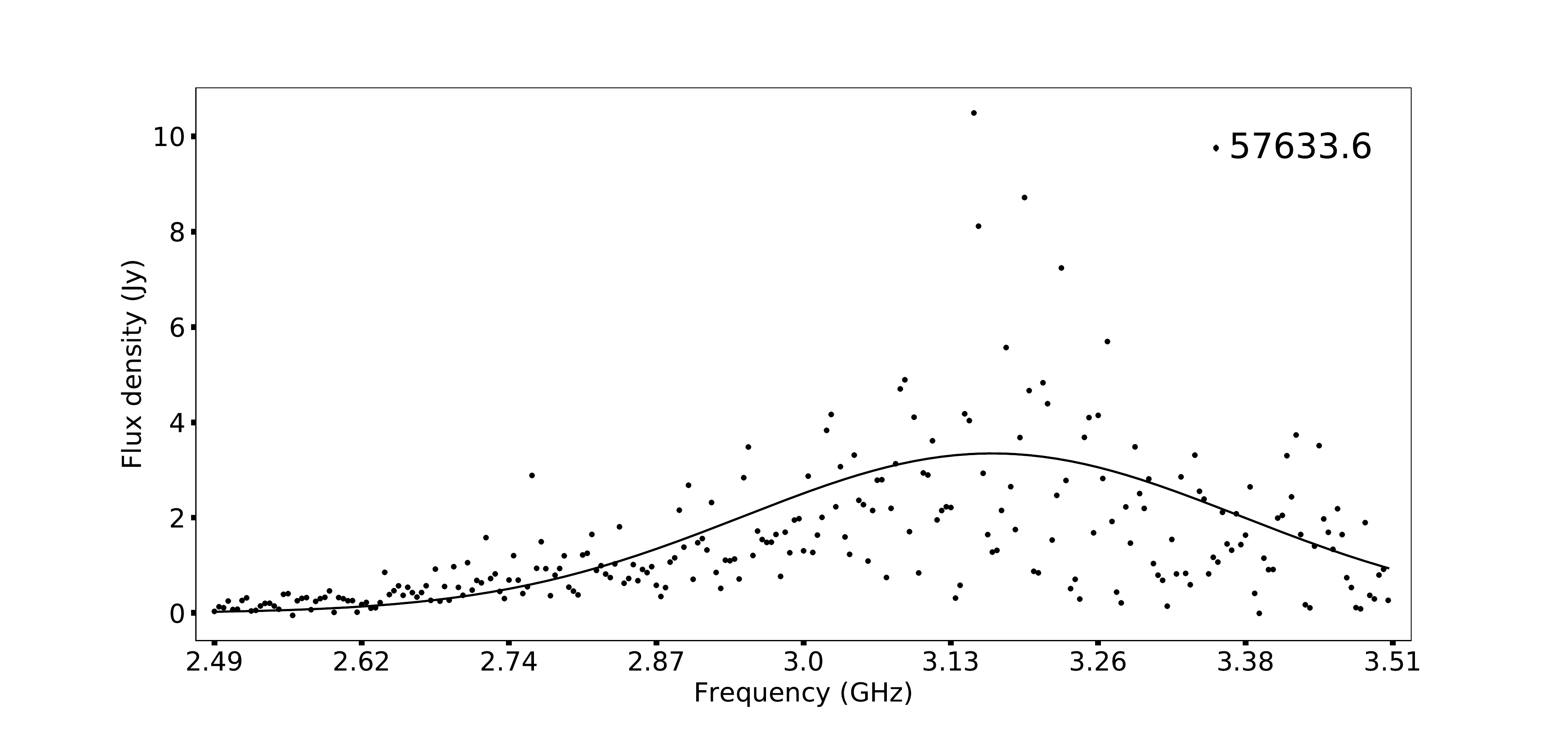}
 \end{minipage}

 \begin{minipage}{2\columnwidth}
  \includegraphics[width=0.5\columnwidth]{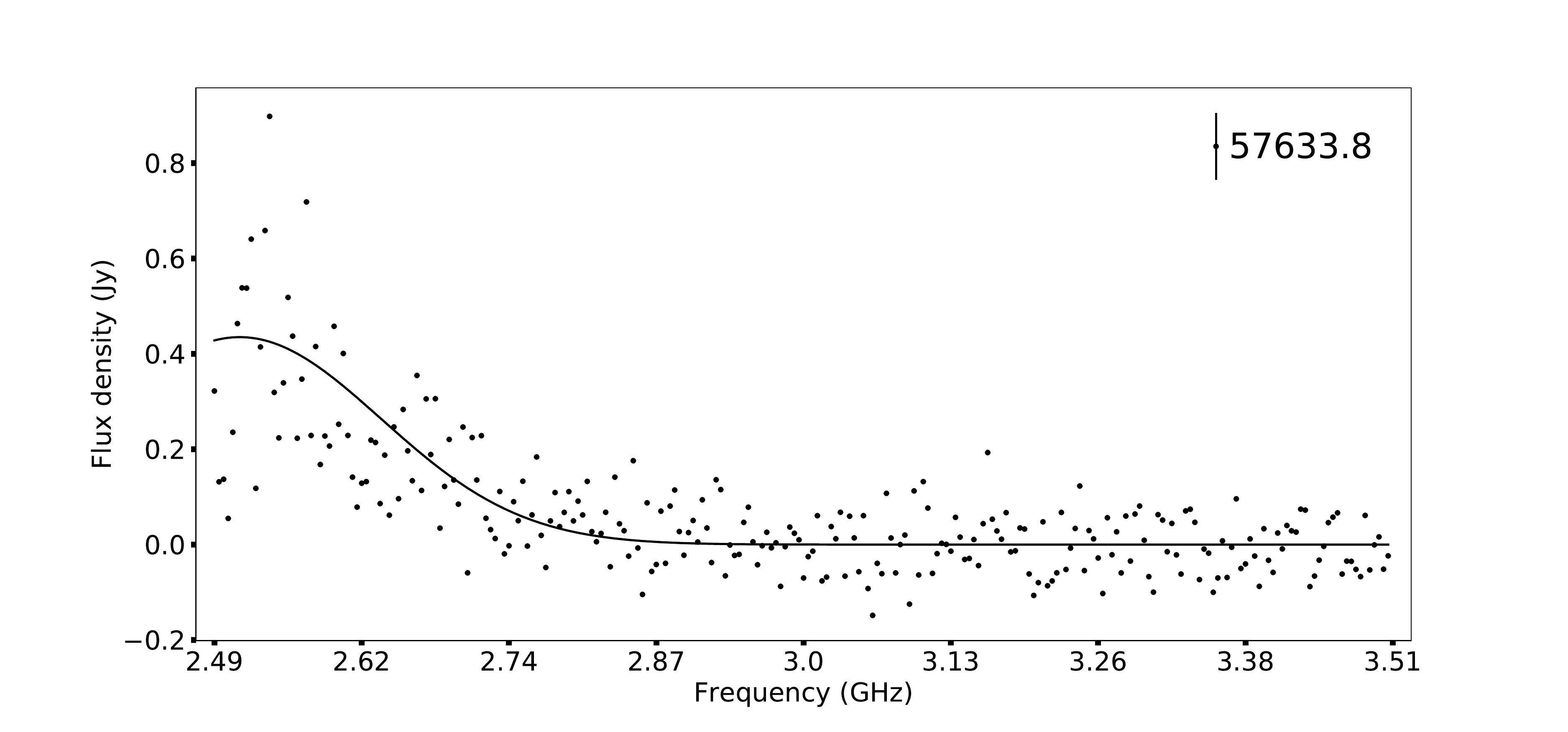}
  \includegraphics[width=0.5\columnwidth]{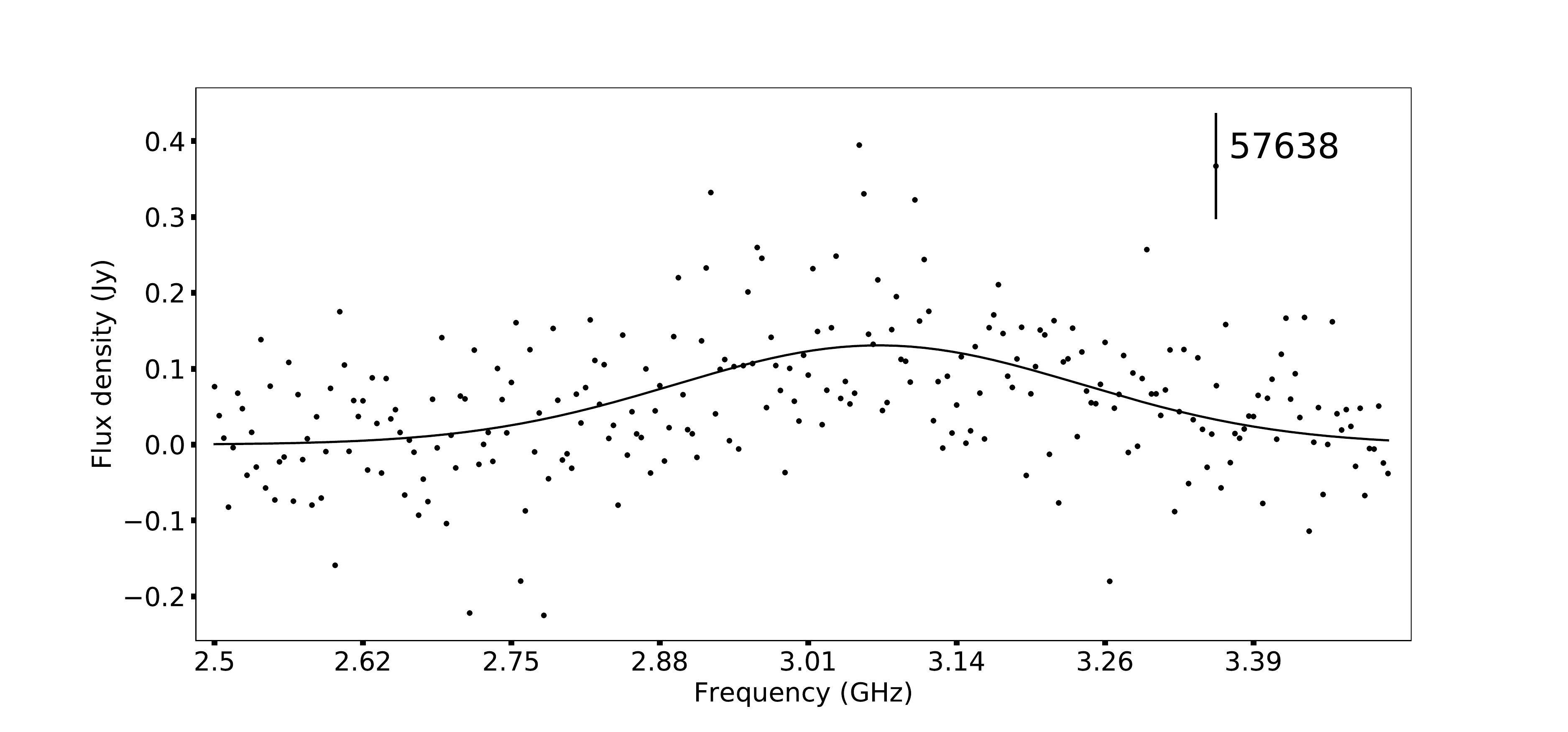}
 \end{minipage}

 \begin{minipage}{2\columnwidth}
  \includegraphics[width=0.5\columnwidth]{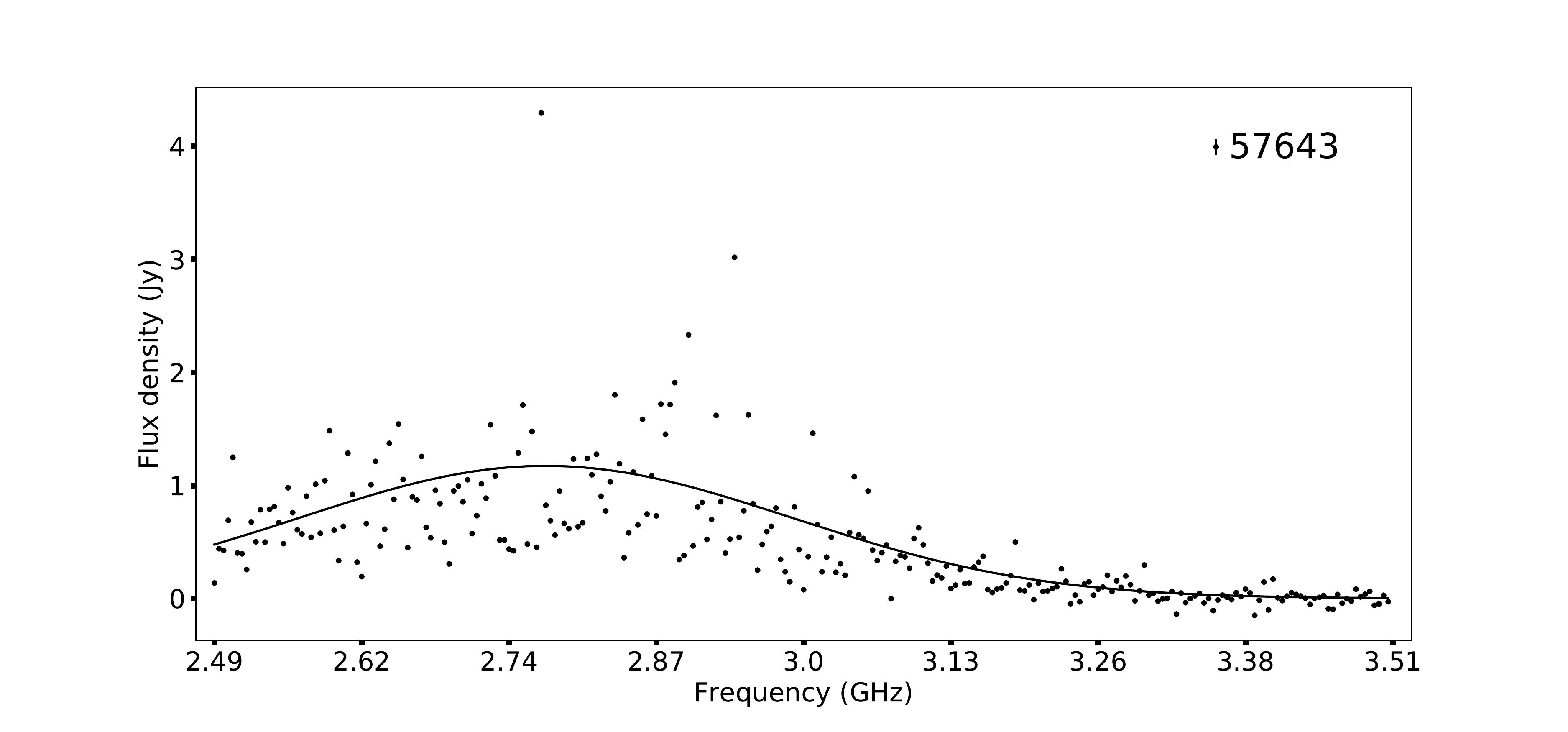}
  \includegraphics[width=0.5\columnwidth]{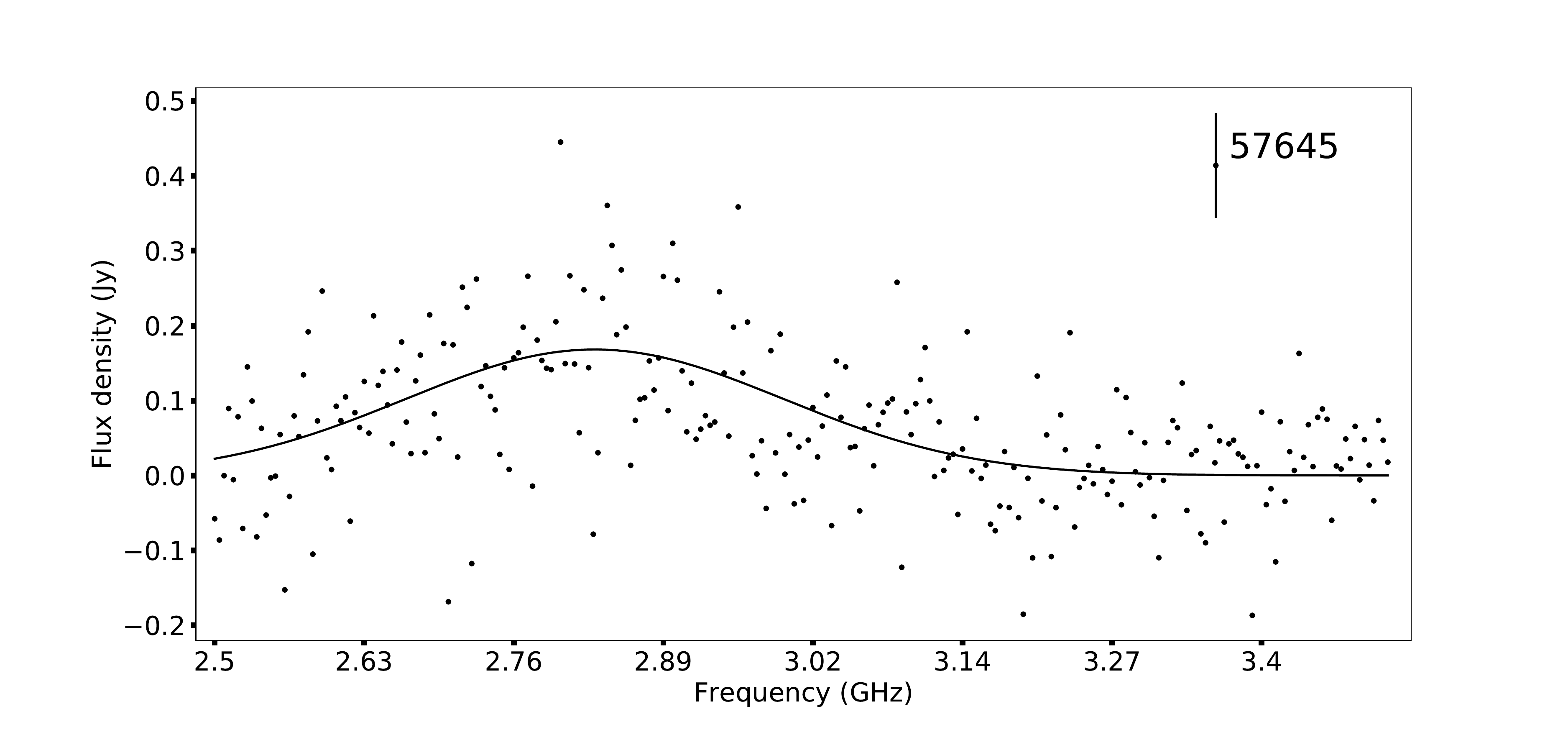}
 \end{minipage}

 \begin{minipage}{2\columnwidth}
  \includegraphics[width=0.5\columnwidth]{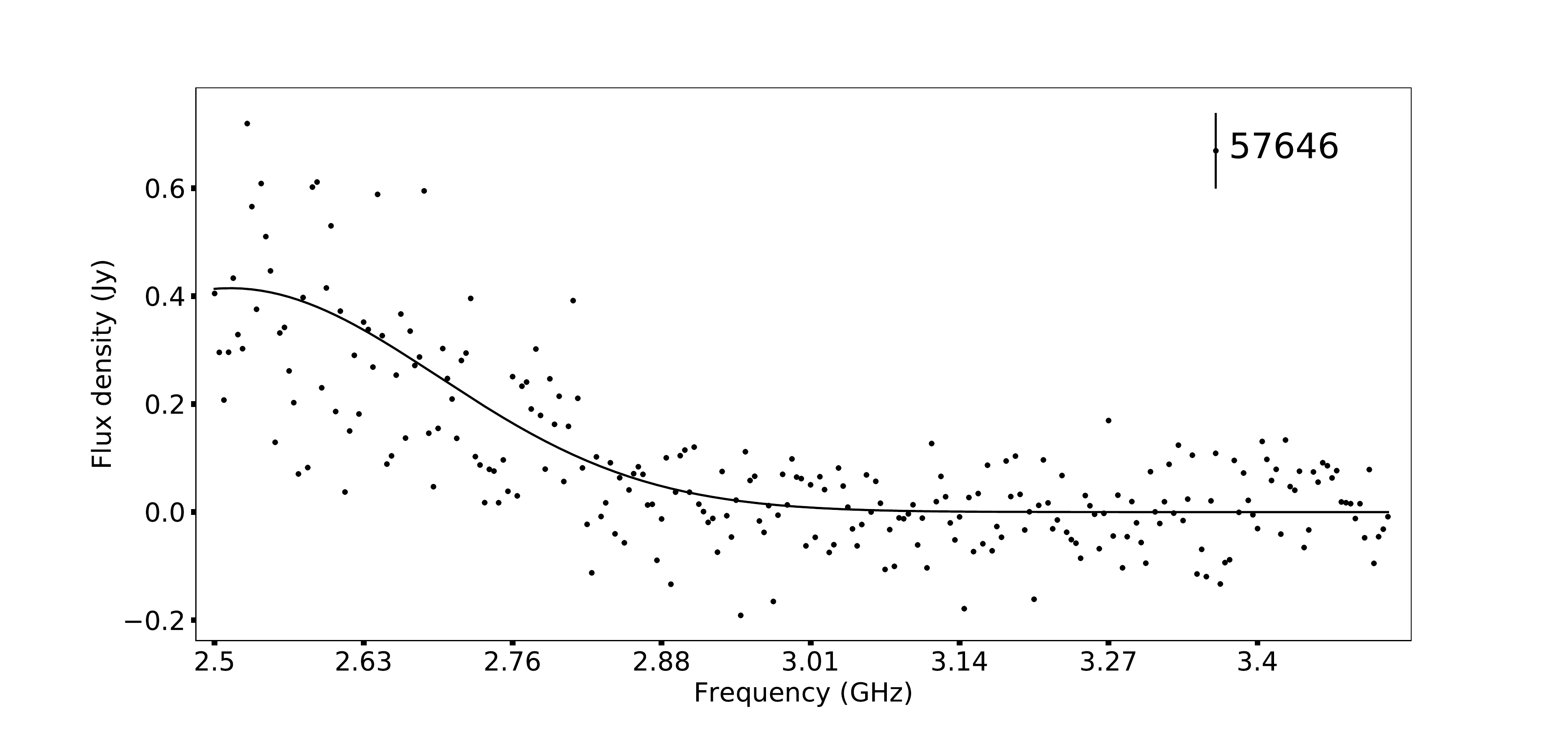}
  \includegraphics[width=0.5\columnwidth]{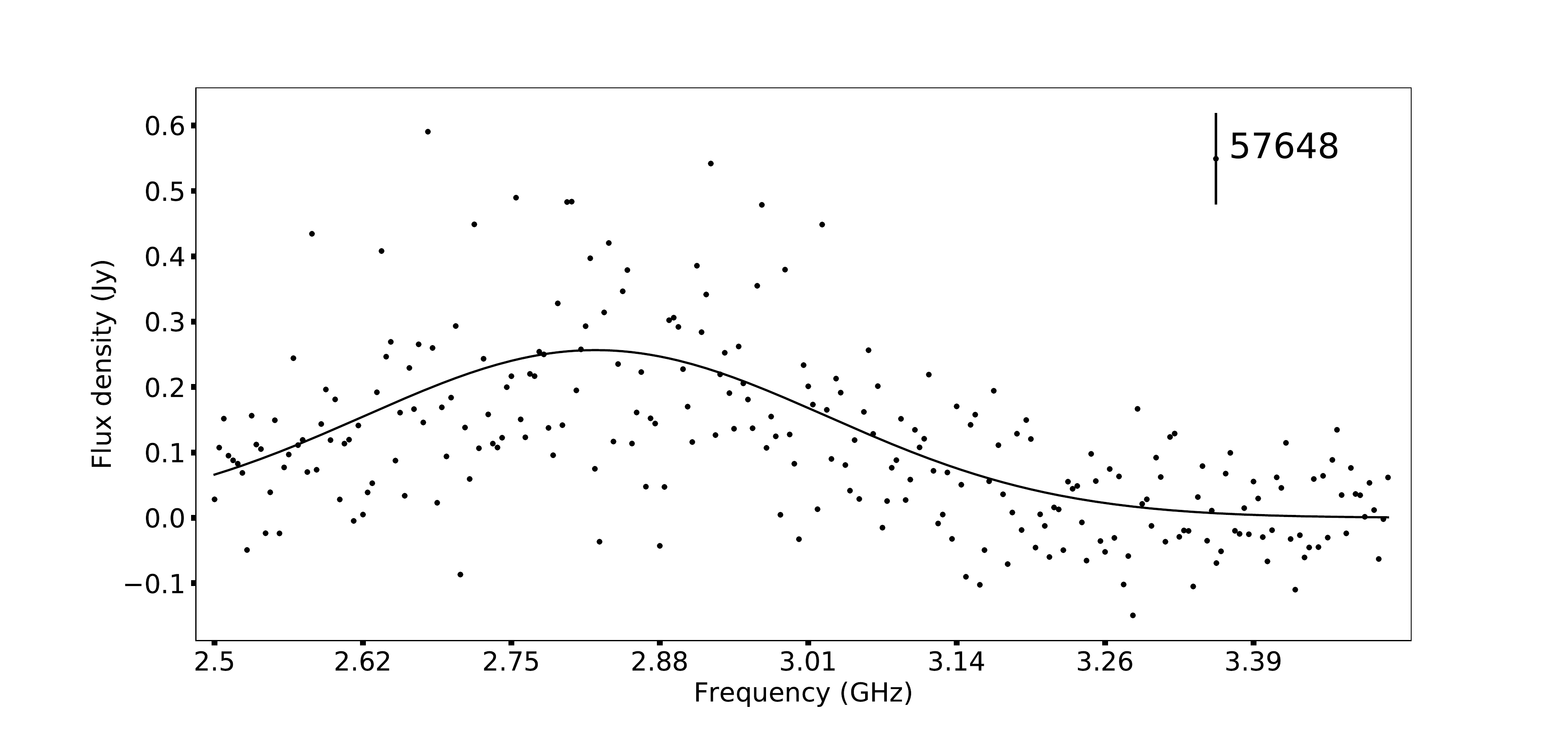}
 \end{minipage}

 \begin{minipage}{2\columnwidth}
  \includegraphics[width=0.5\columnwidth]{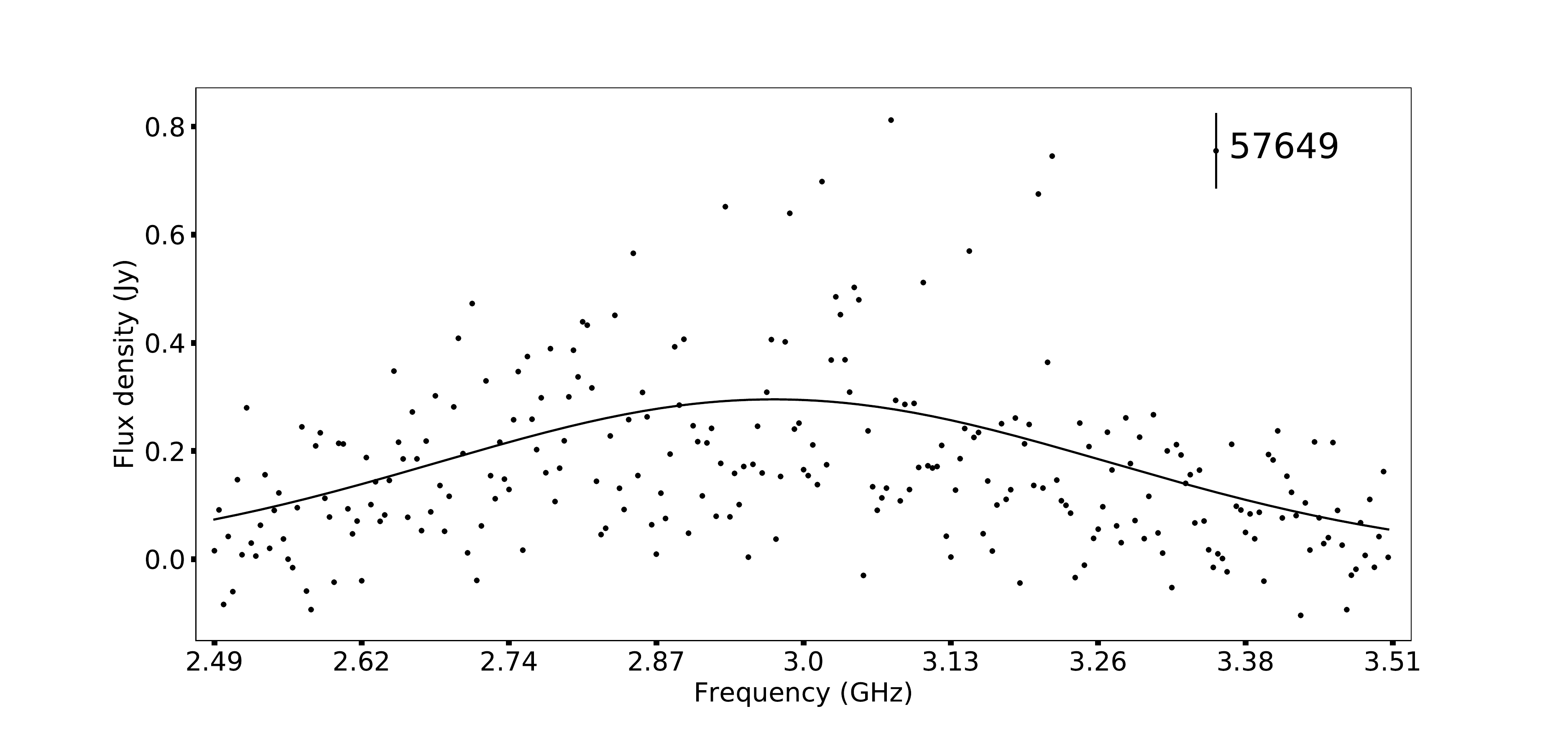}
  \includegraphics[width=0.5\columnwidth]{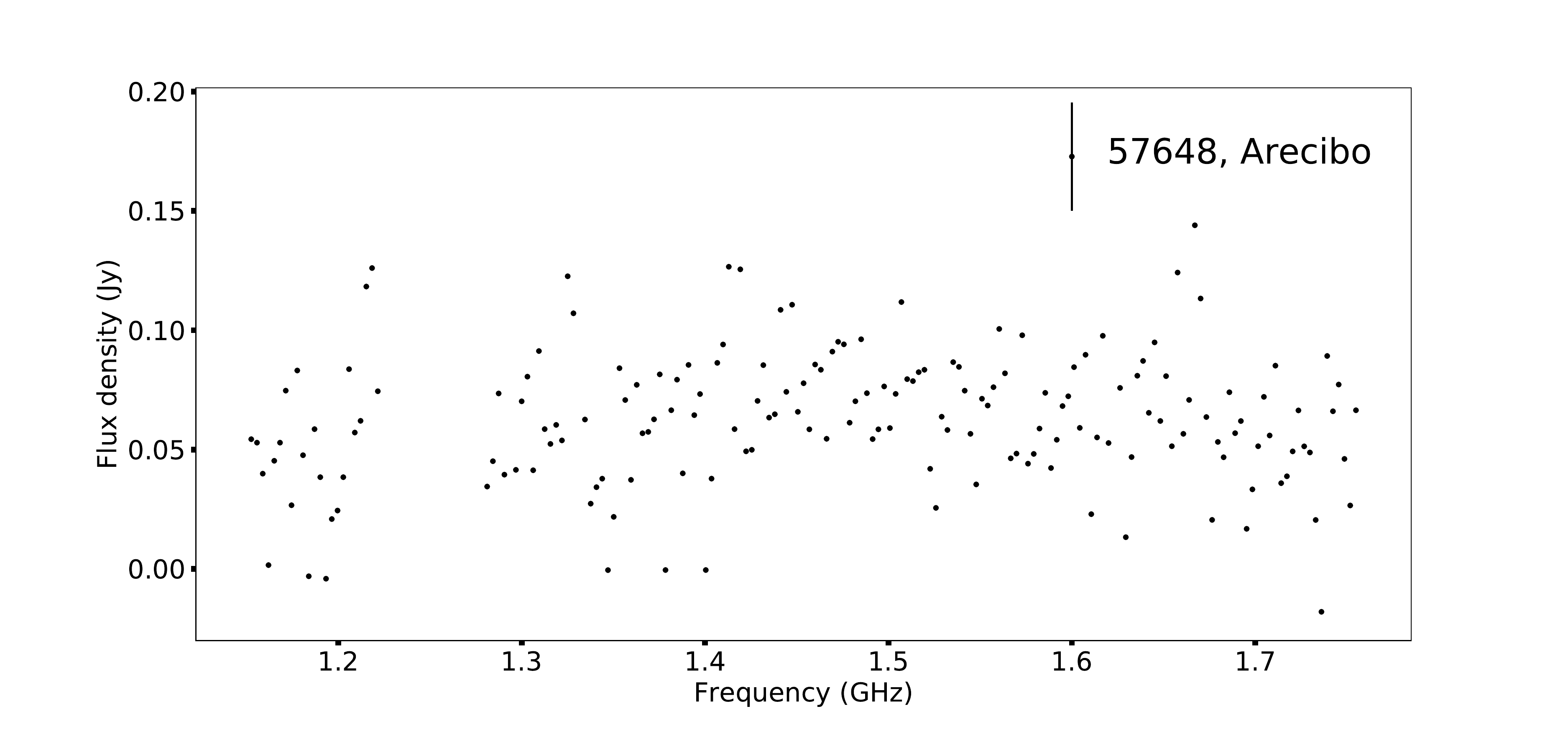}
 \end{minipage}
\caption{Ten panels show spectra for nine VLA bursts and one Arecibo burst (bottom right). The VLA spectra are drawn from flux-calibrated, dedispersed 5~ms integrations. The Arecibo spectrum is drawn from a 2.7~ms window with 3.125~MHz channels and the flux scale is estimated using the radiometer equation. The solid line shown with VLA spectra is a best-fit Gaussian model found through modeling; no comparable modeling is done for the Arecibo spectrum. The typical VLA flux density error per channel is 70~mJy, which is shown at the top right of each panel next to the MJD label.
\label{fig:spec}}
\end{center}
\end{figure*}

Figure \ref{fig:spec} shows that the Stokes I spectra are generally characterized by a broad, Gaussian shape with inter-channel modulation as high as 100\%. Diffractive scintillation from the Milky Way has a typical bandwidth of 7~MHz at a reference frequency of 3~GHz along this line of sight \citep{2002astro.ph..7156C}, which is similar to the channel size of 4~MHz. The burst energy is calculated from the flux density in 5~ms and integrating over the optimal Gaussian spectral model.

The dynamic radio spectra (time versus frequency) were modeled using a Markov chain Monte Carlo (MCMC) method. We used the Goodman and Weare affine invariant sampler \citep{goodman2010ensemble} as implemented in \emph{emcee} \citep{2013PASP..125..306F}. The dynamic spectra were modeled prior to dispersion correction. The spectral structure was modeled as a broad Gaussian envelope as:
\begin{equation}
G(\nu) = \rm{A} \exp{\left[-\frac{1}{2} \left(\frac{\nu - \nu_0}{\sigma}\right)^2\right]}
\end{equation}
\noindent where ``A'' is in mJy per 5~ms integration, $\nu_0$\ is the Gaussian center frequency, and $\sigma$\ is its width. The arrival time was modeled with a cold plasma dispersion law with the arrival time in units of integrations of:
\begin{equation}
\rm{i} (\nu) = \rm{i0} + 4.1488\times10^{-3}\ \rm{DM} \left(\nu^{-2} - \nu_0^{-2}\right)/(5 \rm{ms})
\end{equation}
\noindent where ``i0'' is the arrival integration at the highest frequency and $\nu$\ is in units of GHz. Finally, the pulses were assumed to have an intrinsically square temporal width (parameter $W_{int}$). These 6 parameters defined the intrinsic signal that was then distributed over a fixed time-frequency observing grid.

We use a log likelihood function $\ln \mathcal{L}=(-1/2)\ \sum_i (S_i - S_{i, \rm{model}})^2/\sigma_s^2$, where $S_i$\ refers to the measured and modeled flux density in a single pixel of the dynamic spectrum and $\sigma_s\approx 70~\mu$Jy is the measured off-burst noise per 5~ms integration and 4~MHz channel. A flat prior was used over all ranges with valid data with the requirement that the integrated signal must have a detection significance higher than 8$\sigma$. The 6-dimensional models were sampled with 100 chains taking 700 steps; we ignored the first 200 steps to properly sample the posterior distributions. 

The parameters for the best model of the dynamic spectrum are given in Table \ref{tab:spec} and the resulting Gaussian model overlaid on Figure \ref{fig:spec}.  Two of the best-fit Gaussians are centered at the boundary of the 3~GHz band, so parameter estimates are actually limits. The other seven best-fit models appear contained by the 1~GHz wide band ($>90$\% of the modeled flux is within the 3~GHz band). In all cases, the typical burst spectrum has a FWHM of approximately 500~MHz. This confirms previous reports based on Arecibo and GBT data \citep{2016Natur.531..202S, 2016arXiv160308880S} with the wider (1~GHz) VLA bandwidth and extends this phenomenon to 3~GHz.

Our approach simultaneously models the spectral and temporal evolution of the burst. A typical \frb\ burst takes $\sim180$~ms or 36 integrations to cross the observing band from 2.5 to 3.5~GHz. The effect of a finite pulse width is visible as the burst moves from one integration to the next, even for widths narrower than the integration time of 5~ms. Table \ref{tab:spec} shows that the typical burst width is measured as $2\pm0.5$~ms (68\% confidence interval); the intra-channel dispersion smearing ranges from 0.4 to 1.1~ms across this band. The brightest burst (on MJD 57633.68) is modeled with a temporal width of $2.05\pm0.02$~ms and its arrival time is measured with a precision of $\sim50~\mu$s. Note that these errors are only accurate to the degree that the model represents the data. One potential bias in the model is that we do not model scintillation effects or frequency-dependent temporal width.

\begin{figure*}[htb]
\begin{center}
\includegraphics[width=2\columnwidth]{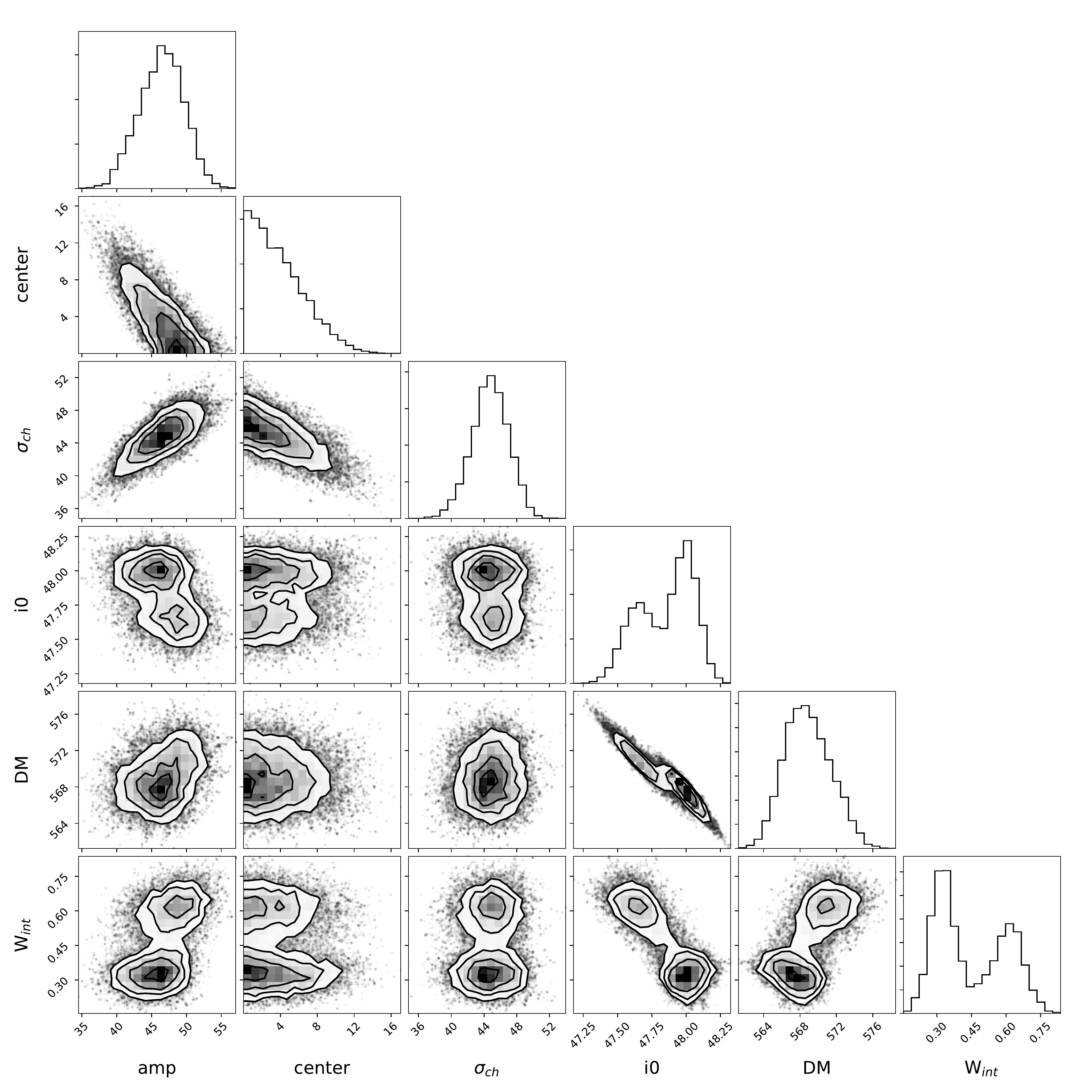}
\caption{A scatterplot matrix that shows the correlation of every pair of parameters in the MCMC run for burst 57646. The equivalent plot for the other bursts show normally-distributed samples, but this burst shows significant structure in the samples. The parameters ``amp'', ``center'', and ``$\sigma_{ch}$'' refer to the Gaussian spectral shape, while ``DM'', ``i0'', and ``W$_{int}$'' refer to the temporal shape. The parameters are given in units of channels (4 MHz) and integrations (5 ms).
\label{fig:corner}}
\end{center}
\end{figure*}

While most bursts are modeled with well-defined parameter probability distributions, one burst is an outlier. Figure \ref{fig:corner} shows the scatter plot matrix \citep{corner} for the MCMC run on burst 57646. Two clusters of samples are identified for this burst: one narrow, low-DM and one wide, high-DM. This shows that the model is not appropriate for this burst and could indicate that the burst can be decomposed into two subbursts.

\subsubsection{Dispersion}

\begin{figure}[htb]
\begin{center}
\includegraphics[width=\columnwidth]{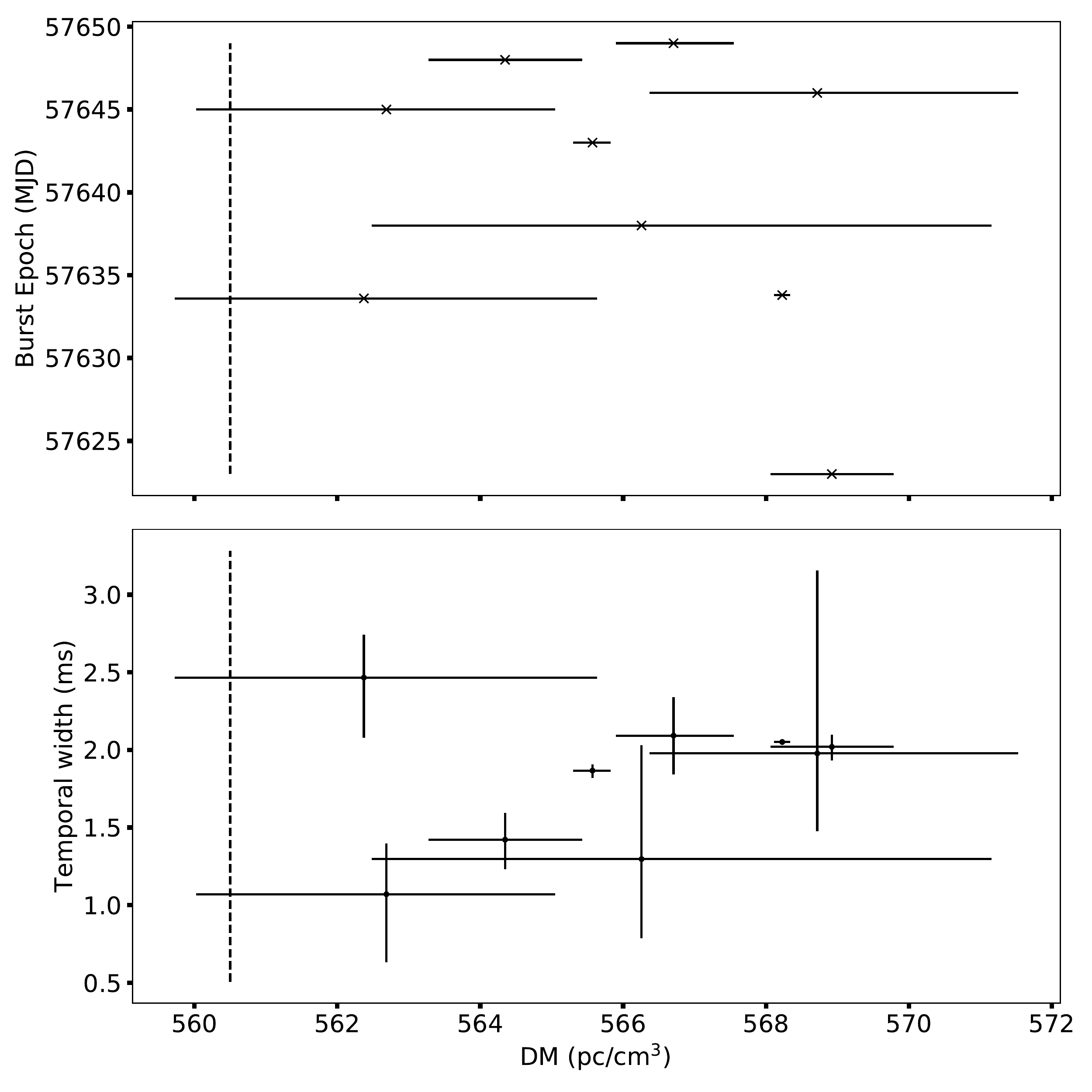}
\caption{(Top) Burst epoch versus DM for nine 3~GHz bursts detected by the VLA. The DM scale for both panels is shown at the bottom. The 68\% confidence interval on DM is shown with a bar and the dashed line shows the best-fit DM$=560.5$\ pc cm$^{-3}$ inferred from Arecibo observations at 1.4~GHz (Hessels et al, in prep). (Bottom) Temporal width versus burst DM for the same nine VLA bursts.
\label{fig:dmdt}}
\end{center}
\end{figure}

The spectrotemporal modeling presented in \S \ref{sec:spec} provides marginalized posterior distributions for burst DM and pulse width. Figure \ref{fig:dmdt} compares the 68\% confidence intervals on the DM for all nine VLA bursts against detection time (top panel) and the modeled burst temporal width (bottom panel). The error weighted mean burst DM is 567.8$\pm$0.1 pc cm$^{-3}$, significantly larger than the long-term average of 560.5 pc cm$^{-3}$ seen by Arecibo during this campaign. Furthermore, several of the 95\% confidence intervals in DM are not consistent with this mean nor each other, which suggests that some burst-specific property can bias the measured DM. The variation in DM observed by the VLA at 3~GHz is similar to that reported for Arecibo observations of \frb\ in \citet{2016Natur.531..202S, 2016arXiv160308880S}.

The bottom panel of Figure \ref{fig:dmdt} compares the DM to the modeled temporal width of the bursts. There is a weak correlation between burst width and apparent DM. It is not clear that whether this correlation is driven by intrinsic structure or unmodeled effects, such as frequency-dependent temporal width. A change in width of $\sim1$~ms correlates with a change in apparent DM of approximately $5$~pc cm$^{-3}$.

\subsubsection{Fine Spectral Structure}

The spectral fits in Figure \ref{fig:spec} have residuals that include single samples that deviate by many standard deviations, particularly in the bursts on MJDs 57623,  57633.6, and 57643. These do not appear to be caused by terrestrial interference. They are most likely spectral variations from diffractive Galactic scintillations that, as noted previously, have characteristic frequency widths slightly larger than the 4~MHz channel bandwidth.  Scintillation variations are multiplicative and have a one-sided exponential probability distribution function that can yield occasional large-amplitude ``scintles'' that are several times the mean scintle amplitude of unity.

The spectral variations could in principle also include contributions from self noise in the source but to be significant, intrinsic fine structure in the burst would have to be on microsecond scales.  

\subsection{Temporal, Energy, and Brightness Distributions of Bursts}
\label{sec:disn}

The burst rate for \frb\ varied dramatically throughout the 2016 observing campaign. In the early-2016 campaign, we observed for 30 hours at 3~GHz and no bursts were detected. In the late-2016 observing campaign, we observed for 27 hours at 3~GHz and detected nine bursts. Overall, the data quality is uniform and high, so the inhomogeneous burst distribution shows that the burst rate was not uniform.

Assuming that the burst detection probability follows a Poisson distribution, the mean VLA, 3~GHz burst rate is $R=0.16\pm0.05\ \rm{hr}^{-1}$ above a fluence of 0.2 Jy ms. The nondetection in the first half of 3~GHz observing limits the FRB rate to $R<0.1$\ hour$^{-1}$\ (95\% confidence limit). The 3~GHz burst rate was much higher during the late-2016 campaign, $R=0.3\pm0.1$\ hour$^{-1}$. This confirms recent analysis of published bursts from \frb\ \citep{2017arXiv170504881O}.

The integrated burst flux density is calculated by integrating the burst spectral model (Table \ref{tab:spec}) in frequency and time. Some VLA burst spectra seem to be contained by the 2.5 to 3.5~GHz band and most of them seem to have Gaussian envelopes that are well modeled by the emission within that band. Assuming that the Gaussian shape defines the full emission window, the integrated 3~GHz flux density can be converted to a total isotropic energy with no further assumptions about the burst spectral properties as:
\begin{equation}
E_{\rm{int}} = (\rm{A} \times 5\ \rm{ms}) ( 2.355 \sigma_{ch}\ \times 4\ \rm{MHz}) 10^{-23} (4\pi L_d^2)\ \rm{erg}
\end{equation}
\noindent where the first term represents the burst fluence, the second term represents the integral of the burst spectrum, 2.355 scales $\sigma$\ to FWHM, and $L_d$\ is the luminosity distance to the source.

\begin{figure}[htb]
\begin{center}
\includegraphics[width=\columnwidth]{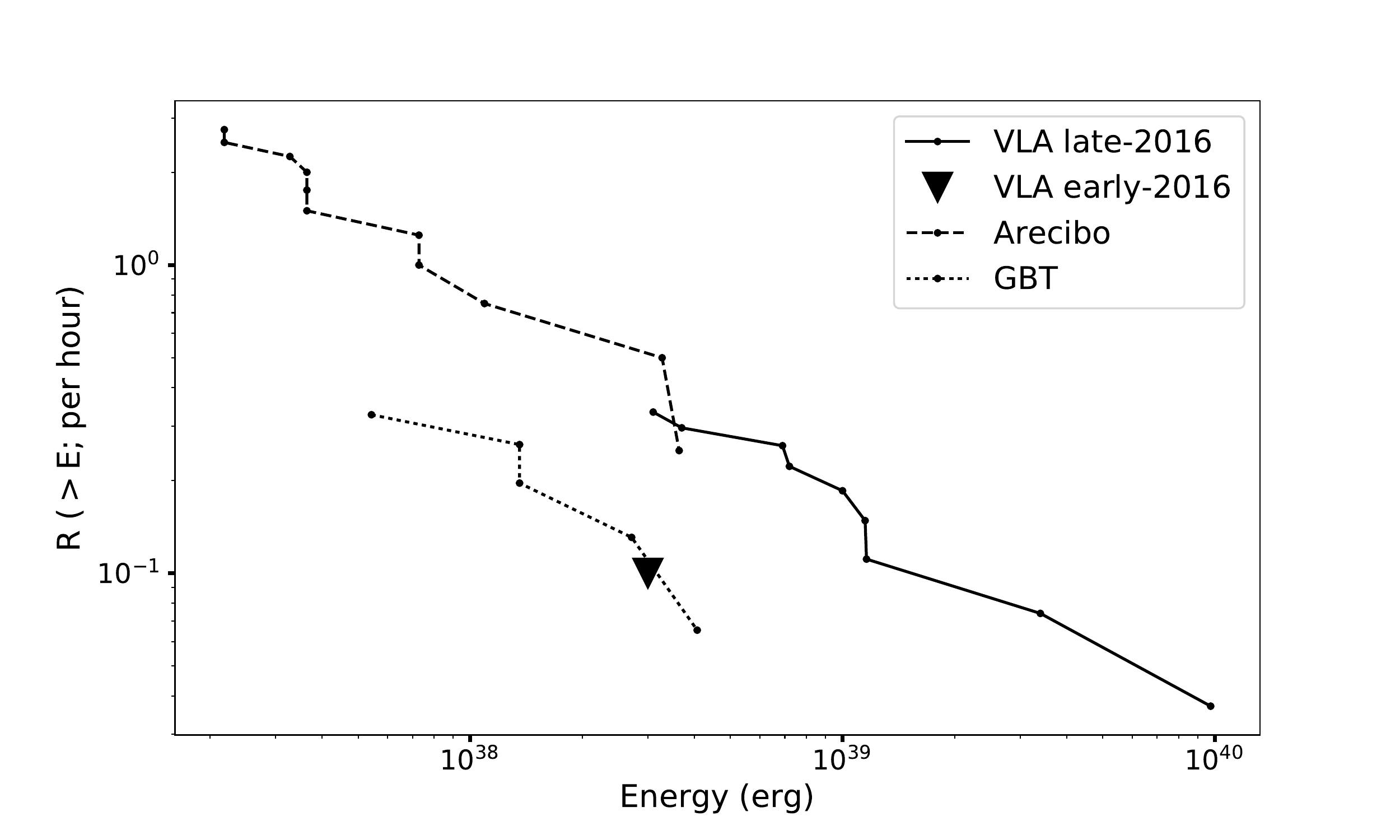}
\caption{The cumulative burst isotropic energy distributions for the VLA, Arecibo, and GBT bursts are shown with dots connected by solid, dashed, and dotted lines, respectively. The Arecibo distribution is derived from the 11 1.4~GHz bursts reported in \citet{2016Natur.531..202S} and the GBT distribution is derived from the 5 1.4~GHz bursts reported in \citet{2016arXiv160308880S}. An upper limit from the VLA nondetection in early 2016 is shown as a triangle. \label{fig:ed}}
\end{center}
\end{figure}

Figure \ref{fig:ed} shows the \frb\ burst energy cumulative distribution as seen by the VLA and calculated from prior observations by Arecibo and the Green Bank Telescope \citep{2016Natur.531..202S, 2016arXiv160308880S}. The latter two energy distributions are scaled from the fluence by assuming that all burst energy is included by the observation. This likely underestimates the burst energy for some bursts, but should not affect the slope of the distribution. The VLA distribution represents a relatively long campaign, so it is sensitive to lower event rates, but is less sensitive than the single-dish campaigns shown (minimum $E_{\rm{int}}\approx3\times10^{38}$\ erg). We also show the rate upper limit (95\% confidence) from the early-2016 VLA campaign to demonstrate that even identical observing campaigns have different detection rates.

We modeled the differential energy distribution, $dN/d\log{E}$, again using a Poisson detection probability with a rate function $\lambda = A E^{\alpha}$. Rather than trying to estimate a completeness limit for each energy distribution, we assume an effective detection limit of 0.9 times the weakest burst detected; the best-fit slope is weakly dependent on this, but general conclusions are robust. We directly sampled the likelihood distribution to estimate a best slope of $\alpha_{\rm{VLA}}=-0.6^{+0.2}_{-0.3}$, $\alpha_{\rm{Arecibo}}=-0.8^{+0.3}_{-0.5}$, and $\alpha_{\rm{GBT}}=-0.8^{+0.4}_{-0.5}$ (68\% confidence interval). Expressed as a power law function in $dN/dE$, the slope index is approximately --1.7. We caution that propagation effects can potentially modulate the intrinsic flux density and significantly affect any interpretation about burst amplitudes. 

\section{discussion}
\label{sec:disc}
\subsection{Burst Spectra}

We present the first simultaneous detection of an FRB with multiple telescopes and over frequencies from 1.2 to 3.5~GHz. The flux density of the Arecibo burst detected on MJD 57648 is an order of magnitude less than that seen by the VLA. At the same time, three other bursts from \frb\ had similar observing coverage but were not detected simultaneously. This is consistent with the spectral structure observed within the 3~GHz VLA band, which is typically limited to a Gaussian envelope with width of roughly 500~MHz. This confirms earlier results at 1.4~GHz \citep{2016Natur.531..202S, 2016arXiv160308880S} with a wider bandwidth and extends the presence of this phenomenon to 3~GHz.

If \frb\ burst spectra are typical of the larger FRB population, then population models need to be generalized beyond the assumption of a spectral power law. Most obviously, \frb\ implies that future multi-telescope searches for bursts are unlikely to simultaneously detect bursts in different bands \citep[c.f.][]{1999ApJ...517..460S}. Second, since burst spectra have limited bands, the burst detection rate at one frequency may not represent that at another. Rate estimates for the FRB population will need to explicitly account for the frequency-dependent rate. Finally, we note that the odds of detecting a burst will improve with bandwidth beyond the gain in sensitivity, since bandwidths wider than the FRB characteristic width are more likley to cover the burst envelope. However, FRB search algorithms will need to be modified to search for bursts with spectral width less than the full bandwidth.

After correcting for barycentric and dispersion delays, the VLA+AO burst spectrum has a residual temporal drift. This drift can be interpreted as an excess DM, but contemporaneous observations at 1.4~GHz measured a significantly lower DM (560.5 pc cm$^{-3}$; Hessels et al, in prep). This excess DM is comparable to the burst-to-burst variation in DM within the whole sample of VLA 3~GHz bursts. This variation had been noted in earlier 1.4~GHz Arecibo observations of \frb\ \citep{2016Natur.531..202S,2016arXiv160308880S}, but the VLA bursts demonstrate that this burst-specific DM is seen simultaneously from 1.4 to 3~GHz.

One interpretation for the burst-dependent DM is that the bursts have a frequency-dependent pulse shape. Pulsar pulse shapes evolve with frequency on GHz frequency scales, presumably due to changes in the beam shape \citep{1988MNRAS.234..477L}. The burst-dependent DM changes are on the order of 1\% of the total DM (equivalent to a delay rate up to 2~ms GHz$^{-1}$), which is much larger than typically observed \citep{2017MNRAS.466.3706L}. Our modeling of the VLA burst dynamic spectra show a weak correlation between larger apparent DM and pulse width. This could be a signature of unresolved spectrotemporal structure in the VLA burst spectra that biases the measured DM.

The burst-specific DM structure could be intrinsic to the emission mechanism or induced during propagation. Radio waves can be modified in a variety of ways (e.g., scintillation, scattering) and the duration of the emission in the source frame is not known \citep{2016arXiv160505890C}. \citet{2017arXiv170306580C} describe a model where plasma lensing near the source of an FRB can magnify radio emission by orders of magnitude. Plasma lensing can also produce multiple burst images separated by time scales from microseconds to tens of milliseconds or longer. These burst images would also have slightly different DMs. Lensing amplifications have a complicated frequency dependence that includes spikes, plateaus, and troughs. The simultaneous detection of burst MJD 57648 with both Arecibo (L band) and the VLA (S band up to 3.2~GHz) requires a focal frequency $\gtrsim 3.2$ GHz and sufficient amplification in both bands.   

\subsection{Energy Distribution}

Our refined analysis shows that \frb\ has isotropic energies as high as $E_{\rm{max}}\approx10^{40}$\ erg. While this emission is coherent and beamed, the apparent energy is larger than Galactic analogues such as the giant pulses from the Crab pulsar \citep[$\sim10^{35}$~erg, $T_{b,\rm{Crab}}\sim10^{41}$~K;][]{2003Natur.422..141H,2014PhRvD..89j3009K}. Radio bursts from \frb\ require either dramatic scaling of known emission processes \citep{2016MNRAS.462..941L, 2016MNRAS.457..232C} and/or strong amplification by propagation effects \citep{2017arXiv170306580C}.

In \S \ref{sec:disn} we compare the energy distribution of the 9 VLA bursts to those reported earlier by Arecibo and the GBT. All three energy distributions can be characterized as a power law in $dN/dE$ with slope $\sim-1.7$. This slope is seen even though the burst rate varies by almost an order of magnitude between campaigns and the observing frequencies cover both 1.4 and 3~GHz bands. This suggests that the slope is related to the underlying physical process, rather than the burst detection rate at any given time. The value of the energy distribution slope is similar to that derived from high-energy bursts from magnetars \citep{2000ApJ...532L.121G, 2011ApJ...739...94S}, which may be examples of self-regulated critical phenomena \citep[slope$=-5/3$;][]{2011SoPh..274...99A}. This is consistent with the idea that magnetar bursts can generate both the millisecond-duration bursts and persistent radio emission from \frb\ \citep{2017arXiv170208644B}.

\subsection{Flux Distribution}

New FRB discoveries have often surprised observers with their remarkable brightness \citep{2007Sci...318..777L, 2016arXiv161105758R}. When modeling the flux distribution as a power law, a uniform, isotropic distribution of sources will have a power law index, $\alpha$, of --1.5. Prior to the measurement of the first GRB redshift, deviations from this Euclidean distribution were used to infer their cosmological distribution \citep[e.g., the $V/V_{\rm{max}}$\ test;][]{1992ApJ...388L..45M, 1995ApJ...453...25F}. Multiple studies have suggested that FRBs have a sub-Euclidean flux distribution \citep[$-0.5<\alpha<-0.9$;][]{2016ApJ...830...75V, 2016arXiv160206099L, 2016arXiv161100458L}. Others have noted that the flux distribution is not well modeled as a power law, such that $\alpha$\ is effectively sensitivity (and telescope) dependent \citep{2016MNRAS.461..984O, 2017arXiv170208040C}.

We now know that \frb\ would be detectable with the VLA out to $z=0.7$, which suggests that cosmological effects are even more important for understanding properites of the FRB population. \citet{2017arXiv170208040C} demonstrate how redshift space distortions, time dilation, and spectral index (``k-correction'') effects can flatten the flux distribution. To this complex scene, we add the fact that \frb\ burst spectra are not defined by a spectral index, which suggests that k-corrections are likely to be very difficult to calculate in practice. If the FRB burst rate is higher at low (rest) radio frequencies, then high-redshift FRBs will have an lower (redshifted) rate that flattens the $dN/dF$\ distribution in a manner similar to a k-correction. However, propagation effects are expected to suppress the FRB rate at sub-GHz frequencies \citep{2017MNRAS.465.2286R, 2017arXiv170107457C} as well. The VLA and Arecibo observed intensively in a coordinated fashion during the late-2016 campaign and a comparison of their relative rates will be presented elsewhere. 

\subsection{Repetition}

The chance of detecting a burst from \frb\ with the VLA changes substantially on day--month timescales. A major outstanding question is whether this time-variable rate is driven by an intrinsic process \citep{2016ApJ...826..226K} or extrinsic (propagation) effects in the host galaxy or intergalactic medium \citep{2017arXiv170306580C}. However, the temporal distribution of VLA detections alone shows that the \frb\ burst rate is significantly correlated on short timescales. That is, detecting a burst implies a higher likelihood of detecting another burst soon thereafter, and not detecting a burst also implies a higher likelihood of not detecting another burst soon thereafter.

The clustering of burst times for the whole FRB population has been modeled as a ``red spectrum'' \citep{2016MNRAS.458L..89C}, while recent work has modeled \frb\ burst times with a modified Poisson distribution \citep{2017arXiv170504881O}. Both temporal models imply that existing observational constraints on repetition are weaker than expected from Poissonian statistics \citep{2015MNRAS.454..457P, 2015ApJ...807...16L}. If the clustered bursts are typical of the FRB population, then wide, shallow surveys are the preferred strategy for blind detection of FRBs. Also, since ``bursts predict bursts'', many short observations are more likely to detect an FRB than a single long observation of the same total length and all new FRB detections should be immediately and intensively followed up.

\subsection{Volumetric Rate of FRB Sources}

%The calculation of a volumetric rate has been used to study classes of optical and high-energy transient \citep{2006ARA&A..44..507W}. 
\citet{2013Sci...341...53T} estimated an isotropic rate of FRB events of $10^{-3}$\ galaxy$^{-1}$\ yr$^{-1}$ by calculating the number of Milky Way-type galaxies out to a distance implied by assming all extragalactic DM originates in the IGM \citep[$z\approx0.9$\ for DM$\approx z\times900\ \rm{pc}\ \rm{cm}^{-3}$;][]{2003ApJ...598L..79I,2004MNRAS.348..999I}. This rate is high \citep[comparable to the rate of core-collapse supernovae;][]{2006Natur.439...45D}, which has been used to argue that FRBs are likely not associated with other classes of transient, such as LGRBs or rare subclasses of SNe \citep{2006ARA&A..44..507W}. However, since this rate depends on telescope sensitivity, it is not appropriate to compare to other estimates based on source counts.

Here, we reevaluate the FRB volumetric rate by assuming \frb\ is a prototype of the class and recasting it as a volumetric rate of FRB sources (i.e., a birth rate). Crudely, the volumetric rate of FRB sources can be defined as $R_{\rm{FRB}} = R_p /(N_r \Omega_b V (z))$, where $R_p$\ is the projected FRB rate, $N_r$\ is the number of bursts per source in a typical lifetime, $\Omega_b$\ is the beaming fraction, and $V(z)$\ is the comoving volume out to redshift $z$. The latest estimates of the projected FRB rate are $2\times10^3\ \rm{sky}^{-1} \rm{day}^{-1}$\ at high Galactic latitudes and flux densities brighter than 1~Jy~ms \citep{2016arXiv161100458L, 2016MNRAS.460L..30C, 2016MNRAS.455.2207R}. There is little constraint on the beaming factor, but Galactic pulsars have beaming solid angle fractions on the order of 10\% \citep{1998MNRAS.298..625T}.

\frb\ has an absolute energy scale that can be used to estimate a horizon scale for the typical FRB survey sensitivity. At the typical Parkes survey parameters \citep[e.g.,][]{2016MNRAS.460L..30C}, a survey with a fluence limit of 1~Jy~ms can detect \frb\ at $z\approx1$ (i.e., for isotropic energy of $\approx10^{40}$~erg). The distance inferred from the largest DMs observed \citep[$\sim1500$\ pc cm$^{-3}$;][]{2016MNRAS.460L..30C} is somewhat higher, but that likely overestimates distance since DM is expected to have significant contributions from the host galaxy and intervening galaxies \citep{OPT, 2014ApJ...780L..33M}.
%For reference, we note that cosmic star formation rate peaks between redshift of 2 and 3\citep{2014ARA&A..52..415M}.

Using an assumed horizon redshift of 1, we estimate a rate of $R_{\rm{FRB}} \approx 5\times10^{-5} N_r^{-1} (0.1/\Omega_b)$\ Mpc$^{-3}$\ yr$^{-1}$. This calculation ignores time dilation, which underestimates the burst rate by a factor of 2 at the redshift horizon. The narrow burst spectral structure seen for \frb\ also suggests that the FRB rate is understated by a factor of a few by ignoring bursts that fall outside the observing band. Finally, we note that this estimate assumes that Galactic latitude and propagation effects do not significantly affect the detectability of FRBs, although scintillation and scattering likely play significant roles \citep{2015MNRAS.451.3278M, 2017arXiv170306580C}.

Despite its many assumptions, this volumetric birth rate is more appropriate for comparison to rates used for other classes of transient. The LGRB and SLSN-I rates are $10^{-7}$\ and $10^{-8}$\ Mpc$^{-3}$\ yr$^{-1}$, respectively \citep{2007ApJ...657L..73G,2012Sci...337..927G}. Comparing the FRB rate to these rates implies that FRB-emitting sources can be generated in LGRBs and SLSN-I if FRBs repeat $5\times10^2$\ and $5\times10^3$\ times\footnote{\citet{2017arXiv170400022N} provide a similar estimate of FRB birth rate and reach a similar conclusion.}, respectively. This assumes isotropic burst energies of $\sim10^{40}$~erg or intrinsic energies of $\sim10^{39} (\Omega_b/0.1)$~erg. \citet{2017arXiv170102370M} use the young magnetar model for FRBs to calculate the maximum number of bursts that can be powered by its magnetic field as $\approx3000 (0.1/\Omega_b)$. This rate comparison suggests that the SLSN-I rate is most consistent with the FRB rate, although there is significant room for adjustment in the models and rate estimate.

%\subsection{Naming Convention}
%The consistency of \frb\ properties (e.g., energy distribution) with the overall population suggests other FRBs likely repeat. If so, the current naming convention for FRBs (analogous to cataclysmic supernovae), will likely become uninformative, especially as large FRB survey projects come online \citep{2014SPIE.9145E..22B}. If all FRBs repeat, then a more useful convention would be based on coordinates. However, there are already two FRBs that have consistent celestial positions and different DMs. Therefore, we suggest an FRB naming convention as Jhhmm+ddDMmmmm.

\section{Conclusions}

The recent precision localization of \frb\ has helped identify a host galaxy, measure its distance, and establish it as a member of a truly new class of astrophysical source. With its cosmological distance firmly established, \frb\ now serves as a new kind of standard by which FRBs are defined. That fact, combined with a rich data set of many bursts, allows us to apply this source to more general questions about the FRB population.

We presented the first multi-telescope detection (Arecibo and VLA) of an FRB. By detecting this burst from 1.2 to 3.5~GHz, we have demonstrated that some bursts have broad spectral structure. However, the energy of that burst was dominated by the VLA 3~GHz observing band and three other VLA bursts are undetected by simultaneous observations at other telescopes. This demonstrates the burst spectra are poorly described by a power law with a single spectral index. We also modelled dynamic spectra within the VLA 3~GHz band and show that most bursts can be characterized by a Gaussian envelope of width $\sim500$\ MHz. This modeling also shows that the apparent DM changes from burst to burst and that DM is biased above the long-term average measured at 1.4~GHz. The nature of this DM change is not known, but could be explained by strong frequency-dependent profile evolution or unresolved spectrotemporal structure in the bursts.

With a characteristic burst spectrum, we can estimate total isotropic radio energy per burst. The cumulative energy distribution is characterized by a power law in $dN/dE$ with slope of $\sim-1.7$. The amplitude of this power law changes significantly between observational campaigns. The stochastic nature of the burst rate suggests that past constraints on FRB repetition are weaker than previously inferred. The relatively narrow spectral structure, flat energy distribution, and variable burst rate of \frb\ suggests that repeated observations, wide bandwidth, and large instantaneous field of view all improve the odds of FRB discovery.

Assuming that \frb\ is representative of the FRB population, we calculate a volumetric birth rate of FRB sources that does not depend on telescope sensitivity. We estimate a volumetric rate of FRB sources of $R_{\rm{FRB}} \approx 5\times10^{-5} (0.1/\Omega_b)$\ Mpc$^{-3}$\ yr$^{-1}$, which is appropriate for \frb-like burst energies of $10^{40}$\ erg that are detectable out to $z=1$. This rate is broadly consistent with models of FRBs from young pulsars or magnetars born in SLSN-I or LGRB, if the typical FRB repeats on the order of $10^3$\ times over its lifetime.

New, arcsecond-scale localizations will be critical to refining the picture presented here and constraining models of FRB origin. \frb\ was localized within hours by a prototype version of \rf\ and an expanded \rf\ system is now under construction. This platform will search a TB/hour data stream in real time in parallel with ongoing VLA observations, potentially detecting and localizing multiple FRBs per year.

\bibliographystyle{yahapj}

\section*{Acknowledgements}
We thank the VLA staff for their support of \rf\ development and Liam Connor for useful feedback. We acknowledge partial support from the Research Corporation for Scientific Advancement (RCSA) for participation in the meeting Fast Radio Bursts: New Probes of Fundamental Physics and Cosmology at the Aspen Center for Physics (February 12-17, 2017).

% observatories
The National Radio Astronomy Observatory is a facility of the National Science Foundation operated under cooperative agreement by Associated Universities, Inc..
Part of this research was carried out at the Jet Propulsion Laboratory, California Institute of Technology, under a contract with the National Aeronautics and Space Administration.
% systems
This research made use of Astropy, a community-developed core Python package for Astronomy (Astropy Collaboration, 2013).
%individuals
CJL is supported by the University of California Office of the President under Lab Fees Research Program Award 237863 and NSF award 1611606. MAM is supported by NSF award 1458952. 
JWTH, CGB, and DM acknowledge support from the European Research Council under the European Union's Seventh Framework Programme (FP/2007-2013) / ERC Grant Agreement nr.\ 337062 (DRAGNET; PI Hessels). JWTH also acknowledges funding from an NWO Vidi fellowship.
SPT acknowledges support from the McGill Astrophysics Fellowship.
KPM's research is supported by the Oxford Centre for Astrophysical Surveys which is funded through the Hintze Family Charitable Foundation. AS gratefully acknowledges support from the European Research Council under grant ERC-2012- StG-307215 LODESTONE. The AMI-LA telescope gratefully acknowledges support from the European Research Council under grant ERC-2012- StG-307215 LODESTONE, the UK Science and Technology Facilities Council (STFC) and the University of Cambridge. 
PS holds a Covington Fellowship at DRAO. 
The LWA1 station is supported by the National Science Foundation under grant 1139974 of the University Radio Observatory program.
VMK acknowledges support from NSERC, CIFAR, the Canada Research Chair Program and from the Lorne Trottier Chair in Astrophysics \& Cosmology.

\facility{EVLA, Arecibo, Effelsberg, LWA, AMI}

\software{\emph{rtpipe}, \emph{realfast}, pwkit \citep{2017ascl.soft04001W}, emcee}

\bibliography{fasttrants.bib}

\begin{thebibliography}{}
\providecommand\natexlab[1]{#1}
\providecommand\JournalTitle[1]{#1}

\bibitem[{{Aschwanden}(2011)}]{2011SoPh..274...99A}
{Aschwanden}, M.~J. 2011,
  \href{http://dx.doi.org/10.1007/s11207-011-9755-0}{\JournalTitle{\solphys},
  274, 99}

\bibitem[{{Beloborodov}(2017)}]{2017arXiv170208644B}
{Beloborodov}, A.~M. 2017, \JournalTitle{ArXiv e-prints},
  \href{http://arxiv.org/abs/1702.08644}{{\sffamily arXiv:1702.08644
  [astro-ph.HE]}}

\bibitem[{{Champion} {et~al.}(2016){Champion}, {Petroff}, {Kramer}, {Keith},
  {Bailes}, {Barr}, {Bates}, {Bhat}, {Burgay}, {Burke-Spolaor}, {Flynn},
  {Jameson}, {Johnston}, {Ng}, {Levin}, {Possenti}, {Stappers}, {van Straten},
  {Thornton}, {Tiburzi}, \& {Lyne}}]{2016MNRAS.460L..30C}
{Champion}, D.~J., {Petroff}, E., {Kramer}, M., {et~al.} 2016,
  \href{http://dx.doi.org/10.1093/mnrasl/slw069}{\JournalTitle{\mnras}, 460,
  L30}

\bibitem[{{Chatterjee} {et~al.}(2017){Chatterjee}, {Law}, {Wharton},
  {Burke-Spolaor}, {Hessels}, {Bower}, {Cordes}, {Tendulkar}, {Bassa},
  {Demorest}, {Butler}, {Seymour}, {Scholz}, {Abruzzo}, {Bogdanov}, {Kaspi},
  {Keimpema}, {Lazio}, {Marcote}, {McLaughlin}, {Paragi}, {Ransom}, {Rupen},
  {Spitler}, \& {van Langevelde}}]{LOC}
{Chatterjee}, S., {Law}, C.~J., {Wharton}, R.~S., {et~al.} 2017,
  \href{http://dx.doi.org/10.1038/nature20797}{\JournalTitle{\nat}, 541, 58}

\bibitem[{{Chawla} {et~al.}(2017){Chawla}, {Kaspi}, {Josephy}, {Rajwade},
  {Lorimer}, {Archibald}, {DeCesar}, {Hessels}, {Kaplan}, {Karako-Argaman},
  {Kondratiev}, {Levin}, {Lynch}, {McLaughlin}, {Ransom}, {Roberts}, {Stairs},
  {Stovall}, {Swiggum}, \& {van Leeuwen}}]{2017arXiv170107457C}
{Chawla}, P., {Kaspi}, V.~M., {Josephy}, A., {et~al.} 2017, \JournalTitle{ArXiv
  e-prints}, \href{http://arxiv.org/abs/1701.07457}{{\sffamily arXiv:1701.07457
  [astro-ph.HE]}}

\bibitem[{{CHIME Scientific Collaboration} {et~al.}(2017){CHIME Scientific
  Collaboration}, {Amiri}, {Bandura}, {Berger}, {Bond}, {Cliche}, {Connor},
  {Deng}, {Denman}, {Dobbs}, {Domagalski}, {Fandino}, {Gilbert}, {Good},
  {Halpern}, {Hanna}, {Hinks}, {Hinshaw}, {H{\"o}fer}, {Hsyu}, {Klages},
  {Landecker}, {Masui}, {Mena-Parra}, {Newburgh}, {Oppermann}, {Pen},
  {Peterson}, {Pinsonneault-Marotte}, {Renard}, {Shaw}, {Siegel}, {Smith},
  {Storer}, {Tretyakov}, {Vanderlinde}, \& {Wiebe}}]{2017arXiv170208040C}
{CHIME Scientific Collaboration}, {Amiri}, M., {Bandura}, K., {et~al.} 2017,
  \JournalTitle{ArXiv e-prints},
  \href{http://arxiv.org/abs/1702.08040}{{\sffamily arXiv:1702.08040
  [astro-ph.HE]}}

\bibitem[{{Connor} {et~al.}(2016{\natexlab{a}}){Connor}, {Pen}, \&
  {Oppermann}}]{2016MNRAS.458L..89C}
{Connor}, L., {Pen}, U.-L., \& {Oppermann}, N. 2016{\natexlab{a}},
  \href{http://dx.doi.org/10.1093/mnrasl/slw026}{\JournalTitle{\mnras}, 458,
  L89}

\bibitem[{{Connor} {et~al.}(2016{\natexlab{b}}){Connor}, {Sievers}, \&
  {Pen}}]{2016MNRAS.458L..19C}
{Connor}, L., {Sievers}, J., \& {Pen}, U.-L. 2016{\natexlab{b}},
  \href{http://dx.doi.org/10.1093/mnrasl/slv124}{\JournalTitle{\mnras}, 458,
  L19}

\bibitem[{{Cordes} \& {Lazio}(2002)}]{2002astro.ph..7156C}
{Cordes}, J.~M., \& {Lazio}, T.~J.~W. 2002, \JournalTitle{ArXiv Astrophysics
  e-prints}, \href{http://arxiv.org/abs/arXiv:astro-ph/0207156}{{\sffamily
  arXiv:astro-ph/0207156}}

\bibitem[{{Cordes} \& {Wasserman}(2016)}]{2016MNRAS.457..232C}
{Cordes}, J.~M., \& {Wasserman}, I. 2016,
  \href{http://dx.doi.org/10.1093/mnras/stv2948}{\JournalTitle{\mnras}, 457,
  232}

\bibitem[{{Cordes} {et~al.}(2017){Cordes}, {Wasserman}, {Hessels}, {Lazio},
  {Chatterjee}, \& {Wharton}}]{2017arXiv170306580C}
{Cordes}, J.~M., {Wasserman}, I., {Hessels}, J.~W.~T., {et~al.} 2017,
  \JournalTitle{ArXiv e-prints},
  \href{http://arxiv.org/abs/1703.06580}{{\sffamily arXiv:1703.06580
  [astro-ph.HE]}}

\bibitem[{{Cordes} {et~al.}(2016){Cordes}, {Wharton}, {Spitler}, {Chatterjee},
  \& {Wasserman}}]{2016arXiv160505890C}
{Cordes}, J.~M., {Wharton}, R.~S., {Spitler}, L.~G., {Chatterjee}, S., \&
  {Wasserman}, I. 2016, \JournalTitle{ArXiv e-prints},
  \href{http://arxiv.org/abs/1605.05890}{{\sffamily arXiv:1605.05890
  [astro-ph.HE]}}

\bibitem[{{Cordes} {et~al.}(2006){Cordes}, {Freire}, {Lorimer}, {Camilo},
  {Champion}, {Nice}, {Ramachandran}, {Hessels}, {Vlemmings}, {van Leeuwen},
  {Ransom}, {Bhat}, {Arzoumanian}, {McLaughlin}, {Kaspi}, {Kasian}, {Deneva},
  {Reid}, {Chatterjee}, {Han}, {Backer}, {Stairs}, {Deshpande}, \&
  {Faucher-Gigu{\`e}re}}]{2006ApJ...637..446C}
{Cordes}, J.~M., {Freire}, P.~C.~C., {Lorimer}, D.~R., {et~al.} 2006,
  \href{http://dx.doi.org/10.1086/498335}{\JournalTitle{\apj}, 637, 446}

\bibitem[{{Diehl} {et~al.}(2006){Diehl}, {Halloin}, {Kretschmer}, {Lichti},
  {Sch{\"o}nfelder}, {Strong}, {von Kienlin}, {Wang}, {Jean}, {Kn{\"o}dlseder},
  {Roques}, {Weidenspointner}, {Schanne}, {Hartmann}, {Winkler}, \&
  {Wunderer}}]{2006Natur.439...45D}
{Diehl}, R., {Halloin}, H., {Kretschmer}, K., {et~al.} 2006,
  \href{http://dx.doi.org/10.1038/nature04364}{\JournalTitle{\nat}, 439, 45}

\bibitem[{{Dokuchaev} \& {Eroshenko}(2017)}]{2017arXiv170102492D}
{Dokuchaev}, V.~I., \& {Eroshenko}, Y.~N. 2017, \JournalTitle{ArXiv e-prints,
  1701.02492}, \href{http://arxiv.org/abs/1701.02492}{{\sffamily
  arXiv:1701.02492 [astro-ph.HE]}}

\bibitem[{{Ellingson} {et~al.}(2013){Ellingson}, {Taylor}, {Craig}, {Hartman},
  {Dowell}, {Wolfe}, {Clarke}, {Hicks}, {Kassim}, {Ray}, {Rickard}, {Schinzel},
  \& {Weiler}}]{2013ITAP...61.2540E}
{Ellingson}, S.~W., {Taylor}, G.~B., {Craig}, J., {et~al.} 2013,
  \href{http://dx.doi.org/10.1109/TAP.2013.2242826}{\JournalTitle{IEEE
  Transactions on Antennas and Propagation}, 61, 2540}

\bibitem[{{Fenimore} \& {Bloom}(1995)}]{1995ApJ...453...25F}
{Fenimore}, E.~E., \& {Bloom}, J.~S. 1995,
  \href{http://dx.doi.org/10.1086/176366}{\JournalTitle{\apj}, 453, 25}

\bibitem[{Foreman-Mackey(2016)}]{corner}
Foreman-Mackey, D. 2016,
  \href{http://dx.doi.org/10.21105/joss.00024}{\JournalTitle{The Journal of
  Open Source Software}, 24}

\bibitem[{{Foreman-Mackey} {et~al.}(2013){Foreman-Mackey}, {Hogg}, {Lang}, \&
  {Goodman}}]{2013PASP..125..306F}
{Foreman-Mackey}, D., {Hogg}, D.~W., {Lang}, D., \& {Goodman}, J. 2013,
  \href{http://dx.doi.org/10.1086/670067}{\JournalTitle{\pasp}, 125, 306}

\bibitem[{{Fuller} \& {Ott}(2015)}]{2015MNRAS.450L..71F}
{Fuller}, J., \& {Ott}, C.~D. 2015,
  \href{http://dx.doi.org/10.1093/mnrasl/slv049}{\JournalTitle{\mnras}, 450,
  L71}

\bibitem[{{Gal-Yam}(2012)}]{2012Sci...337..927G}
{Gal-Yam}, A. 2012,
  \href{http://dx.doi.org/10.1126/science.1203601}{\JournalTitle{Science}, 337,
  927}

\bibitem[{Goodman \& Weare(2010)}]{goodman2010ensemble}
Goodman, J., \& Weare, J. 2010, \JournalTitle{Communications in applied
  mathematics and computational science}, 5, 65

\bibitem[{{G{\"o}{\v g}{\"u}{\c s}} {et~al.}(2000){G{\"o}{\v g}{\"u}{\c s}},
  {Woods}, {Kouveliotou}, {van Paradijs}, {Briggs}, {Duncan}, \&
  {Thompson}}]{2000ApJ...532L.121G}
{G{\"o}{\v g}{\"u}{\c s}}, E., {Woods}, P.~M., {Kouveliotou}, C., {et~al.}
  2000, \href{http://dx.doi.org/10.1086/312583}{\JournalTitle{\apjl}, 532,
  L121}

\bibitem[{{Guetta} \& {Della Valle}(2007)}]{2007ApJ...657L..73G}
{Guetta}, D., \& {Della Valle}, M. 2007,
  \href{http://dx.doi.org/10.1086/511417}{\JournalTitle{\apjl}, 657, L73}

\bibitem[{{Hankins} {et~al.}(2003){Hankins}, {Kern}, {Weatherall}, \&
  {Eilek}}]{2003Natur.422..141H}
{Hankins}, T.~H., {Kern}, J.~S., {Weatherall}, J.~C., \& {Eilek}, J.~A. 2003,
  \JournalTitle{\nat}, 422, 141

\bibitem[{{Inoue}(2004)}]{2004MNRAS.348..999I}
{Inoue}, S. 2004,
  \href{http://dx.doi.org/10.1111/j.1365-2966.2004.07359.x}{\JournalTitle{\mnras},
  348, 999}

\bibitem[{{Ioka}(2003)}]{2003ApJ...598L..79I}
{Ioka}, K. 2003, \href{http://dx.doi.org/10.1086/380598}{\JournalTitle{\apjl},
  598, L79}

\bibitem[{{Kashiyama} \& {Murase}(2017)}]{2017arXiv170104815K}
{Kashiyama}, K., \& {Murase}, K. 2017, \JournalTitle{ArXiv e-prints,
  1701.04815}, \href{http://arxiv.org/abs/1701.04815}{{\sffamily
  arXiv:1701.04815 [astro-ph.HE]}}

\bibitem[{{Katz}(2014)}]{2014PhRvD..89j3009K}
{Katz}, J.~I. 2014,
  \href{http://dx.doi.org/10.1103/PhysRevD.89.103009}{\JournalTitle{\prd}, 89,
  103009}

\bibitem[{{Katz}(2016)}]{2016ApJ...826..226K}
---. 2016,
  \href{http://dx.doi.org/10.3847/0004-637X/826/2/226}{\JournalTitle{\apj},
  826, 226}

\bibitem[{{Kulkarni} {et~al.}(2014){Kulkarni}, {Ofek}, {Neill}, {Zheng}, \&
  {Juric}}]{2014ApJ...797...70K}
{Kulkarni}, S.~R., {Ofek}, E.~O., {Neill}, J.~D., {Zheng}, Z., \& {Juric}, M.
  2014,
  \href{http://dx.doi.org/10.1088/0004-637X/797/1/70}{\JournalTitle{\apj}, 797,
  70}

\bibitem[{{Law} {et~al.}(2017){Law}, {Bower}, {Burke-Spolaor}, {Butler},
  {Paul}, {Lazio}, \& {Rupen}}]{2017AAS...22933002L}
{Law}, C.~J., {Bower}, G.~C., {Burke-Spolaor}, S., {et~al.} 2017, in American
  Astronomical Society Meeting Abstracts, Vol. 229, American Astronomical
  Society Meeting Abstracts, 330.02

\bibitem[{{Law} {et~al.}(2015){Law}, {Bower}, {Burke-Spolaor}, {Butler},
  {Lawrence}, {Lazio}, {Mattmann}, {Rupen}, {Siemion}, \&
  {VanderWiel}}]{2015ApJ...807...16L}
{Law}, C.~J., {Bower}, G.~C., {Burke-Spolaor}, S., {et~al.} 2015,
  \href{http://dx.doi.org/10.1088/0004-637X/807/1/16}{\JournalTitle{\apj}, 807,
  16}

\bibitem[{{Lawrence} {et~al.}(2016){Lawrence}, {Vander Wiel}, {Law}, {Burke
  Spolaor}, \& {Bower}}]{2016arXiv161100458L}
{Lawrence}, E., {Vander Wiel}, S., {Law}, C.~J., {Burke Spolaor}, S., \&
  {Bower}, G.~C. 2016, \JournalTitle{ArXiv e-prints},
  \href{http://arxiv.org/abs/1611.00458}{{\sffamily arXiv:1611.00458
  [astro-ph.HE]}}

\bibitem[{{Lazarus} {et~al.}(2015){Lazarus}, {Brazier}, {Hessels},
  {Karako-Argaman}, {Kaspi}, {Lynch}, {Madsen}, {Patel}, {Ransom}, {Scholz},
  {Swiggum}, {Zhu}, {Allen}, {Bogdanov}, {Camilo}, {Cardoso}, {Chatterjee},
  {Cordes}, {Crawford}, {Deneva}, {Ferdman}, {Freire}, {Jenet}, {Knispel},
  {Lee}, {van Leeuwen}, {Lorimer}, {Lyne}, {McLaughlin}, {Siemens}, {Spitler},
  {Stairs}, {Stovall}, \& {Venkataraman}}]{2015ApJ...812...81L}
{Lazarus}, P., {Brazier}, A., {Hessels}, J.~W.~T., {et~al.} 2015,
  \href{http://dx.doi.org/10.1088/0004-637X/812/1/81}{\JournalTitle{\apj}, 812,
  81}

\bibitem[{{Lentati} {et~al.}(2017){Lentati}, {Kerr}, {Dai}, {Hobson},
  {Shannon}, {Hobbs}, {Bailes}, {Bhat}, {Burke-Spolaor}, {Coles}, {Dempsey},
  {Lasky}, {Levin}, {Manchester}, {Os{\l}owski}, {Ravi}, {Reardon}, {Rosado},
  {Spiewak}, {van Straten}, {Toomey}, {Wang}, {Wen}, {You}, \&
  {Zhu}}]{2017MNRAS.466.3706L}
{Lentati}, L., {Kerr}, M., {Dai}, S., {et~al.} 2017,
  \href{http://dx.doi.org/10.1093/mnras/stw3359}{\JournalTitle{\mnras}, 466,
  3706}

\bibitem[{{Li} {et~al.}(2016){Li}, {Huang}, {Zhang}, {Li}, \&
  {Li}}]{2016arXiv160206099L}
{Li}, L., {Huang}, Y., {Zhang}, Z., {Li}, D., \& {Li}, B. 2016,
  \JournalTitle{ArXiv e-prints},
  \href{http://arxiv.org/abs/1602.06099}{{\sffamily arXiv:1602.06099
  [astro-ph.HE]}}

\bibitem[{{Lorimer} {et~al.}(2007){Lorimer}, {Bailes}, {McLaughlin},
  {Narkevic}, \& {Crawford}}]{2007Sci...318..777L}
{Lorimer}, D.~R., {Bailes}, M., {McLaughlin}, M.~A., {Narkevic}, D.~J., \&
  {Crawford}, F. 2007,
  \href{http://dx.doi.org/10.1126/science.1147532}{\JournalTitle{Science}, 318,
  777}

\bibitem[{{Lunnan} {et~al.}(2014){Lunnan}, {Chornock}, {Berger}, {Laskar},
  {Fong}, {Rest}, {Sanders}, {Challis}, {Drout}, {Foley}, {Huber}, {Kirshner},
  {Leibler}, {Marion}, {McCrum}, {Milisavljevic}, {Narayan}, {Scolnic},
  {Smartt}, {Smith}, {Soderberg}, {Tonry}, {Burgett}, {Chambers}, {Flewelling},
  {Hodapp}, {Kaiser}, {Magnier}, {Price}, \& {Wainscoat}}]{2014ApJ...787..138L}
{Lunnan}, R., {Chornock}, R., {Berger}, E., {et~al.} 2014,
  \href{http://dx.doi.org/10.1088/0004-637X/787/2/138}{\JournalTitle{\apj},
  787, 138}

\bibitem[{{Lyne} \& {Manchester}(1988)}]{1988MNRAS.234..477L}
{Lyne}, A.~G., \& {Manchester}, R.~N. 1988,
  \href{http://dx.doi.org/10.1093/mnras/234.3.477}{\JournalTitle{\mnras}, 234,
  477}

\bibitem[{{Lyutikov} {et~al.}(2016){Lyutikov}, {Burzawa}, \&
  {Popov}}]{2016MNRAS.462..941L}
{Lyutikov}, M., {Burzawa}, L., \& {Popov}, S.~B. 2016,
  \href{http://dx.doi.org/10.1093/mnras/stw1669}{\JournalTitle{\mnras}, 462,
  941}

\bibitem[{{Macquart} \& {Johnston}(2015)}]{2015MNRAS.451.3278M}
{Macquart}, J.-P., \& {Johnston}, S. 2015,
  \href{http://dx.doi.org/10.1093/mnras/stv1184}{\JournalTitle{\mnras}, 451,
  3278}

\bibitem[{{Mao} \& {Paczynski}(1992)}]{1992ApJ...388L..45M}
{Mao}, S., \& {Paczynski}, B. 1992,
  \href{http://dx.doi.org/10.1086/186326}{\JournalTitle{\apjl}, 388, L45}

\bibitem[{{Marcote} {et~al.}(2017){Marcote}, {Paragi}, {Hessels}, {Keimpema},
  {van Langevelde}, {Huang}, {Bassa}, {Bogdanov}, {Bower}, {Burke-Spolaor},
  {Butler}, {Campbell}, {Chatterjee}, {Cordes}, {Demorest}, {Garrett}, {Ghosh},
  {Kaspi}, {Law}, {Lazio}, {McLaughlin}, {Ransom}, {Salter}, {Scholz},
  {Seymour}, {Siemion}, {Spitler}, {Tendulkar}, \& {Wharton}}]{EVN}
{Marcote}, B., {Paragi}, Z., {Hessels}, J.~W.~T., {et~al.} 2017,
  \href{http://dx.doi.org/10.3847/2041-8213/834/2/L8}{\JournalTitle{\apjl},
  834, L8}

\bibitem[{{McQuinn}(2014)}]{2014ApJ...780L..33M}
{McQuinn}, M. 2014,
  \href{http://dx.doi.org/10.1088/2041-8205/780/2/L33}{\JournalTitle{\apjl},
  780, L33}

\bibitem[{{Metzger} {et~al.}(2017){Metzger}, {Berger}, \&
  {Margalit}}]{2017arXiv170102370M}
{Metzger}, B.~D., {Berger}, E., \& {Margalit}, B. 2017, \JournalTitle{ArXiv
  e-prints,1701.02370}, \href{http://arxiv.org/abs/1701.02370}{{\sffamily
  arXiv:1701.02370 [astro-ph.HE]}}

\bibitem[{{Modjaz} {et~al.}(2008){Modjaz}, {Kewley}, {Kirshner}, {Stanek},
  {Challis}, {Garnavich}, {Greene}, {Kelly}, \& {Prieto}}]{2008AJ....135.1136M}
{Modjaz}, M., {Kewley}, L., {Kirshner}, R.~P., {et~al.} 2008,
  \href{http://dx.doi.org/10.1088/0004-6256/135/4/1136}{\JournalTitle{\aj},
  135, 1136}

\bibitem[{{Nicholl} {et~al.}(2017){Nicholl}, {Williams}, {Berger}, {Villar},
  {Alexander}, {Eftekhari}, \& {Metzger}}]{2017arXiv170400022N}
{Nicholl}, M., {Williams}, P.~K.~G., {Berger}, E., {et~al.} 2017,
  \JournalTitle{ArXiv e-prints},
  \href{http://arxiv.org/abs/1704.00022}{{\sffamily arXiv:1704.00022
  [astro-ph.HE]}}

\bibitem[{{Opperman} \& {Pen}(2017)}]{2017arXiv170504881O}
{Opperman}, N., \& {Pen}, U.-L. 2017, \JournalTitle{ArXiv e-prints},
  \href{http://arxiv.org/abs/1705.04881}{{\sffamily arXiv:1705.04881
  [astro-ph.HE]}}

\bibitem[{{Oppermann} {et~al.}(2016){Oppermann}, {Connor}, \&
  {Pen}}]{2016MNRAS.461..984O}
{Oppermann}, N., {Connor}, L.~D., \& {Pen}, U.-L. 2016,
  \href{http://dx.doi.org/10.1093/mnras/stw1401}{\JournalTitle{\mnras}, 461,
  984}

\bibitem[{{Perrott} {et~al.}(2013){Perrott}, {Scaife}, {Green}, {Davies},
  {Franzen}, {Grainge}, {Hobson}, {Hurley-Walker}, {Lasenby}, {Olamaie},
  {Pooley}, {Rodr{\'{\i}}guez-Gonz{\'a}lvez}, {Rumsey}, {Saunders}, {Schammel},
  {Scott}, {Shimwell}, {Titterington}, {Waldram}, \& {AMI
  Consortium}}]{2013MNRAS.429.3330P}
{Perrott}, Y.~C., {Scaife}, A.~M.~M., {Green}, D.~A., {et~al.} 2013,
  \href{http://dx.doi.org/10.1093/mnras/sts589}{\JournalTitle{\mnras}, 429,
  3330}

\bibitem[{{Petroff} {et~al.}(2015){Petroff}, {Johnston}, {Keane}, {van
  Straten}, {Bailes}, {Barr}, {Barsdell}, {Burke-Spolaor}, {Caleb}, {Champion},
  {Flynn}, {Jameson}, {Kramer}, {Ng}, {Possenti}, \&
  {Stappers}}]{2015MNRAS.454..457P}
{Petroff}, E., {Johnston}, S., {Keane}, E.~F., {et~al.} 2015,
  \href{http://dx.doi.org/10.1093/mnras/stv1953}{\JournalTitle{\mnras}, 454,
  457}

\bibitem[{{Petroff} {et~al.}(2016){Petroff}, {Barr}, {Jameson}, {Keane},
  {Bailes}, {Kramer}, {Morello}, {Tabbara}, \& {van
  Straten}}]{2016PASA...33...45P}
{Petroff}, E., {Barr}, E.~D., {Jameson}, A., {et~al.} 2016,
  \href{http://dx.doi.org/10.1017/pasa.2016.35}{\JournalTitle{\pasa}, 33, e045}

\bibitem[{{Planck Collaboration} {et~al.}(2016){Planck Collaboration}, {Ade},
  {Aghanim}, {Arnaud}, {Ashdown}, {Aumont}, {Baccigalupi}, {Banday},
  {Barreiro}, {Bartlett}, \& et~al.}]{2016A&A...594A..13P}
{Planck Collaboration}, {Ade}, P.~A.~R., {Aghanim}, N., {et~al.} 2016,
  \href{http://dx.doi.org/10.1051/0004-6361/201525830}{\JournalTitle{\aap},
  594, A13}

\bibitem[{{Popov} \& {Pshirkov}(2016)}]{2016MNRAS.462L..16P}
{Popov}, S.~B., \& {Pshirkov}, M.~S. 2016,
  \href{http://dx.doi.org/10.1093/mnrasl/slw118}{\JournalTitle{\mnras}, 462,
  L16}

\bibitem[{{Rajwade} \& {Lorimer}(2017)}]{2017MNRAS.465.2286R}
{Rajwade}, K.~M., \& {Lorimer}, D.~R. 2017,
  \href{http://dx.doi.org/10.1093/mnras/stw2914}{\JournalTitle{\mnras}, 465,
  2286}

\bibitem[{{Rane} {et~al.}(2016){Rane}, {Lorimer}, {Bates}, {McMann},
  {McLaughlin}, \& {Rajwade}}]{2016MNRAS.455.2207R}
{Rane}, A., {Lorimer}, D.~R., {Bates}, S.~D., {et~al.} 2016,
  \href{http://dx.doi.org/10.1093/mnras/stv2404}{\JournalTitle{\mnras}, 455,
  2207}

\bibitem[{{Ransom}(2001)}]{2001PhDT.......123R}
{Ransom}, S.~M. 2001, PhD thesis, Harvard University

\bibitem[{{Ravi} {et~al.}(2016){Ravi}, {Shannon}, {Bailes}, {Bannister},
  {Bhandari}, {Bhat}, {Burke-Spolaor}, {Caleb}, {Flynn}, {Jameson}, {Johnston},
  {Keane}, {Kerr}, {Tiburzi}, {Tuntsov}, \& {Vedantham}}]{2016arXiv161105758R}
{Ravi}, V., {Shannon}, R.~M., {Bailes}, M., {et~al.} 2016, \JournalTitle{ArXiv
  e-prints}, \href{http://arxiv.org/abs/1611.05758}{{\sffamily arXiv:1611.05758
  [astro-ph.HE]}}

\bibitem[{{Sallmen} {et~al.}(1999){Sallmen}, {Backer}, {Hankins}, {Moffett}, \&
  {Lundgren}}]{1999ApJ...517..460S}
{Sallmen}, S., {Backer}, D.~C., {Hankins}, T.~H., {Moffett}, D., \& {Lundgren},
  S. 1999, \href{http://dx.doi.org/10.1086/307183}{\JournalTitle{\apj}, 517,
  460}

\bibitem[{{Scholz} \& {Kaspi}(2011)}]{2011ApJ...739...94S}
{Scholz}, P., \& {Kaspi}, V.~M. 2011,
  \href{http://dx.doi.org/10.1088/0004-637X/739/2/94}{\JournalTitle{\apj}, 739,
  94}

\bibitem[{{Scholz} {et~al.}(2016){Scholz}, {Spitler}, {Hessels}, {Chatterjee},
  {Cordes}, {Kaspi}, {Wharton}, {Bassa}, {Bogdanov}, {Camilo}, {Crawford},
  {Deneva}, {van Leeuwen}, {Lynch}, {Madsen}, {McLaughlin}, {Mickaliger},
  {Parent}, {Patel}, {Ransom}, {Seymour}, {Stairs}, {Stappers}, \&
  {Tendulkar}}]{2016arXiv160308880S}
{Scholz}, P., {Spitler}, L.~G., {Hessels}, J.~W.~T., {et~al.} 2016,
  \JournalTitle{ArXiv e-prints},
  \href{http://arxiv.org/abs/1603.08880}{{\sffamily arXiv:1603.08880
  [astro-ph.HE]}}

\bibitem[{{Spitler} {et~al.}(2014){Spitler}, {Cordes}, {Hessels}, {Lorimer},
  {McLaughlin}, {Chatterjee}, {Crawford}, {Deneva}, {Kaspi}, {Wharton},
  {Allen}, {Bogdanov}, {Brazier}, {Camilo}, {Freire}, {Jenet},
  {Karako-Argaman}, {Knispel}, {Lazarus}, {Lee}, {van Leeuwen}, {Lynch},
  {Ransom}, {Scholz}, {Siemens}, {Stairs}, {Stovall}, {Swiggum},
  {Venkataraman}, {Zhu}, {Aulbert}, \& {Fehrmann}}]{2014ApJ...790..101S}
{Spitler}, L.~G., {Cordes}, J.~M., {Hessels}, J.~W.~T., {et~al.} 2014,
  \href{http://dx.doi.org/10.1088/0004-637X/790/2/101}{\JournalTitle{\apj},
  790, 101}

\bibitem[{{Spitler} {et~al.}(2016){Spitler}, {Scholz}, {Hessels}, {Bogdanov},
  {Brazier}, {Camilo}, {Chatterjee}, {Cordes}, {Crawford}, {Deneva}, {Ferdman},
  {Freire}, {Kaspi}, {Lazarus}, {Lynch}, {Madsen}, {McLaughlin}, {Patel},
  {Ransom}, {Seymour}, {Stairs}, {Stappers}, {van Leeuwen}, \&
  {Zhu}}]{2016Natur.531..202S}
{Spitler}, L.~G., {Scholz}, P., {Hessels}, J.~W.~T., {et~al.} 2016,
  \href{http://dx.doi.org/10.1038/nature17168}{\JournalTitle{\nat}, 531, 202}

\bibitem[{{Tauris} \& {Manchester}(1998)}]{1998MNRAS.298..625T}
{Tauris}, T.~M., \& {Manchester}, R.~N. 1998,
  \href{http://dx.doi.org/10.1046/j.1365-8711.1998.01369.x}{\JournalTitle{\mnras},
  298, 625}

\bibitem[{{Tendulkar} {et~al.}(2017){Tendulkar}, {Bassa}, {Cordes}, {Bower},
  {Law}, {Chatterjee}, {Adams}, {Bogdanov}, {Burke-Spolaor}, {Butler},
  {Demorest}, {Hessels}, {Kaspi}, {Lazio}, {Maddox}, {Marcote}, {McLaughlin},
  {Paragi}, {Ransom}, {Scholz}, {Seymour}, {Spitler}, {van Langevelde}, \&
  {Wharton}}]{OPT}
{Tendulkar}, S.~P., {Bassa}, C.~G., {Cordes}, J.~M., {et~al.} 2017,
  \href{http://dx.doi.org/10.3847/2041-8213/834/2/L7}{\JournalTitle{\apjl},
  834, L7}

\bibitem[{{Thompson}(2017)}]{2017arXiv170300393T}
{Thompson}, C. 2017, \JournalTitle{ArXiv e-prints},
  \href{http://arxiv.org/abs/1703.00393}{{\sffamily arXiv:1703.00393
  [astro-ph.HE]}}

\bibitem[{{Thornton} {et~al.}(2013){Thornton}, {Stappers}, {Bailes},
  {Barsdell}, {Bates}, {Bhat}, {Burgay}, {Burke-Spolaor}, {Champion}, {Coster},
  {D'Amico}, {Jameson}, {Johnston}, {Keith}, {Kramer}, {Levin}, {Milia}, {Ng},
  {Possenti}, \& {van Straten}}]{2013Sci...341...53T}
{Thornton}, D., {Stappers}, B., {Bailes}, M., {et~al.} 2013,
  \href{http://dx.doi.org/10.1126/science.1236789}{\JournalTitle{Science}, 341,
  53}

\bibitem[{{Vedantham} {et~al.}(2016){Vedantham}, {Ravi}, {Hallinan}, \&
  {Shannon}}]{2016ApJ...830...75V}
{Vedantham}, H.~K., {Ravi}, V., {Hallinan}, G., \& {Shannon}, R.~M. 2016,
  \href{http://dx.doi.org/10.3847/0004-637X/830/2/75}{\JournalTitle{\apj}, 830,
  75}

\bibitem[{{Williams} {et~al.}(2017){Williams}, {Clavel}, {Newton}, \&
  {Ryzhkov}}]{2017ascl.soft04001W}
{Williams}, P.~K.~G., {Clavel}, M., {Newton}, E., \& {Ryzhkov}, D. 2017,
  {pwkit: Astronomical utilities in Python}, Astrophysics Source Code Library,
  \href{http://arxiv.org/abs/1704.001}{{\sffamily ascl:1704.001}}

\bibitem[{{Woosley} \& {Bloom}(2006)}]{2006ARA&A..44..507W}
{Woosley}, S.~E., \& {Bloom}, J.~S. 2006,
  \href{http://dx.doi.org/10.1146/annurev.astro.43.072103.150558}{\JournalTitle{\araa},
  44, 507}

\bibitem[{{Zhang}(2017)}]{2017arXiv170104094Z}
{Zhang}, B. 2017, \JournalTitle{ArXiv e-prints, 1701.04094},
  \href{http://arxiv.org/abs/1701.04094}{{\sffamily arXiv:1701.04094
  [astro-ph.HE]}}

\bibitem[{{Zwart} {et~al.}(2008){Zwart}, {Barker}, {Biddulph}, {Bly}, {Boysen},
  {Brown}, {Clementson}, {Crofts}, {Culverhouse}, {Czeres}, {Dace}, {Davies},
  {D'Alessandro}, {Doherty}, {Duggan}, {Ely}, {Felvus}, {Feroz}, {Flynn},
  {Franzen}, {Geisb{\"u}sch}, {G{\'e}nova-Santos}, {Grainge}, {Grainger},
  {Hammett}, {Hills}, {Hobson}, {Holler}, {Hurley-Walker}, {Jilley}, {Jones},
  {Kaneko}, {Kneissl}, {Lancaster}, {Lasenby}, {Marshall}, {Newton}, {Norris},
  {Northrop}, {Odell}, {Petencin}, {Pober}, {Pooley}, {Pospieszalski}, {Quy},
  {Rodr{\'{\i}}guez-Gonz{\'a}lvez}, {Saunders}, {Scaife}, {Schofield}, {Scott},
  {Shaw}, {Shimwell}, {Smith}, {Taylor}, {Titterington}, {Veli{\'c}},
  {Waldram}, {West}, {Wood}, {Yassin}, \& {AMI
  Consortium}}]{2008MNRAS.391.1545Z}
{Zwart}, J.~T.~L., {Barker}, R.~W., {Biddulph}, P., {et~al.} 2008,
  \href{http://dx.doi.org/10.1111/j.1365-2966.2008.13953.x}{\JournalTitle{\mnras},
  391, 1545}

\end{thebibliography}

\end{document}